\documentclass[aps, prx, superscriptaddress,reprint]{revtex4-2}

\usepackage{braket,amsmath,amssymb,graphicx,comment,xcolor,float}
\usepackage[T1]{fontenc}
\usepackage{times}
\usepackage{xcolor}
\usepackage{hyperref}
\usepackage{siunitx}
\usepackage{ulem}
\usepackage{graphicx}
\usepackage{mathtools}
\usepackage{amssymb}
\usepackage{amsthm}
\usepackage{thmtools}
\newcommand{\abs}[1]{\left\lvert #1 \right\rvert}
\newcommand{\avg}[1]{\left\langle #1 \right\rangle}
\usepackage{amsmath}

\newcommand{\refsub}[2]{\hyperref[#1]{\ref*{#1}#2}}

\DeclareSIUnit\cps{cps}
\DeclareSIUnit\photons{photons}
\usepackage{overpic}

\begin{document} 

\title{Error-Detected Quantum Operations with Neutral Atoms Mediated by an Optical Cavity} 

\author{Brandon Grinkemeyer}
\altaffiliation{These authors contributed equally to this work}
\affiliation{Department of Physics, Harvard University, Cambridge, MA 02138, USA}
\author{Elmer Guardado-Sanchez}
\altaffiliation{These authors contributed equally to this work}
\affiliation{Department of Physics, Harvard University, Cambridge, MA 02138, USA}
\author{Ivana Dimitrova}
\altaffiliation{These authors contributed equally to this work}
\affiliation{Department of Physics, Harvard University, Cambridge, MA 02138, USA}
\author{Danilo Shchepanovich}
\affiliation{Department of Physics, Harvard University, Cambridge, MA 02138, USA}
\author{G. Eirini Mandopoulou}
\affiliation{Department of Physics, Harvard University, Cambridge, MA 02138, USA}
\author{Johannes Borregaard}
\affiliation{Department of Physics, Harvard University, Cambridge, MA 02138, USA}
\author{Vladan Vuleti\ifmmode \acute{c}\else \'{c}\fi{}}
\affiliation{Department of Physics and Research Laboratory of Electronics, Massachusetts Institute of Technology, Cambridge, MA 02139, USA}
\author{Mikhail D. Lukin}
\affiliation{Department of Physics, Harvard University, Cambridge, MA 02138, USA}

\begin{abstract} 
Neutral atom quantum processors are a promising platform for large-scale quantum computing. Integrating them with an optical cavity enables fast nondestructive qubit readout and access to fast remote entanglement generation for quantum networking. Here, we introduce a platform for coupling single atoms in optical tweezers to a Fabry-Perot Fiber Cavity. Leveraging the strong atom-cavity coupling, we demonstrate fast qubit state readout with 99.960$^{+14}_{-24}\SI{}{\percent}$ fidelity and two methods for cavity-mediated entanglement generation with integrated error detection. First, we use cavity-carving to generate a Bell state with 91(4)$\%$ fidelity and a 32(1)$\%$ success rate. Second, we perform a cavity-mediated gate with a deterministic entanglement fidelity of 52.5(18)$\%$, increased to 76(2)$\%$ with error detection. The new capabilities enabled by this platform pave the way towards modular quantum computing and networking.
\end{abstract}

\date{\today}
\maketitle

\section*{Introduction}

Neutral atom arrays recently emerged as a promising platform for large scale quantum information systems, enabling quantum algorithms with multiple logical qubits \cite{bluvstein2024logical} and novel approaches to quantum metrology and clocks \cite{madjarov2019atomic, young2020half, cao2024multi, shaw2024multi}. These recent advancements are facilitated by the implementation of high-fidelity two-qubit gates and the use of coherent transport for non-local connectivity and reconfigurable architecture \cite{evered2023high, ma2023high, scholl2023erasure, bluvstein2022quantum}. While quantum processors with over 10,000 physical qubits appear within the reach \cite{manetsch2024tweezer}, 
further scaling may benefit from a modular approach, in which quantum computation is distributed across quantum processors connected by fast high-fidelity quantum network channels \cite{monroe14modular,young2022architecture, huie2021multiplexed, ocola2024nanophotonic,li24highrate, sinclair2024fault}. Such an approach requires integration of atom arrays with  optical cavities, which provides direct coupling to photons in a well-defined mode that can be used for fast high-fidelity remote entanglement distribution \cite{reiserer2015cavity, daiss2021quantum}. Integration of individually-controlled atoms in optical tweezers with an optical cavity has only recently been realized experimentally \cite{dhordjevic2021entanglement, yan23superradiant, liu23array, hartung24array}.

Here, we demonstrate a platform combining individual control and transport of atoms in optical tweezers with efficient coupling to individual optical photons enabled by a Fabry-Perot fiber cavity (FPFC). These cavities offer some of the largest cooperativities to date (C$\sim$100) \cite{hunger2010fiber, gehr2010cavity}, while also providing optical access for optical tweezers. This approach can be directly integrated with a reconfigurable architecture where a neutral atom array is placed above the cavity and select atoms are coherently transported in and out of the cavity mode \cite{dhordjevic2021entanglement, ocola2024nanophotonic}. This further enhances the capabilities of neutral atom quantum computer, enabling an efficient distributed  processor. We carry out several experiments to demonstrate the capabilities of this platform. Specifically, we demonstrate that the high cooperativity enables fast high-fidelity qubit state readout, which can be important for accelerating quantum error-correction algorithms. Moreover, we show that error detection can be naturally integrated within this approach to enable robust entanglement generation. Note that unlike previous demonstrations of cavity-mediated entanglement that relied on photon detection, our protocol heralds success based on the atomic state, which significantly increases the success rate of entanglement generation, thereby improving the efficiency of quantum networking scheme that relies on probabilistic Bell measurements \cite{langenfeld2021quantum}. This method can be used for fast and more complex quantum network operations, such as entanglement swapping, purification, and fusion, which can enable the development of an efficient distributed quantum computer \cite{thomas2024fusion, singh24modular, dur24modular}.

\begin{figure*}
\centering
\noindent\makebox[\textwidth]{\includegraphics[width=5.7in]{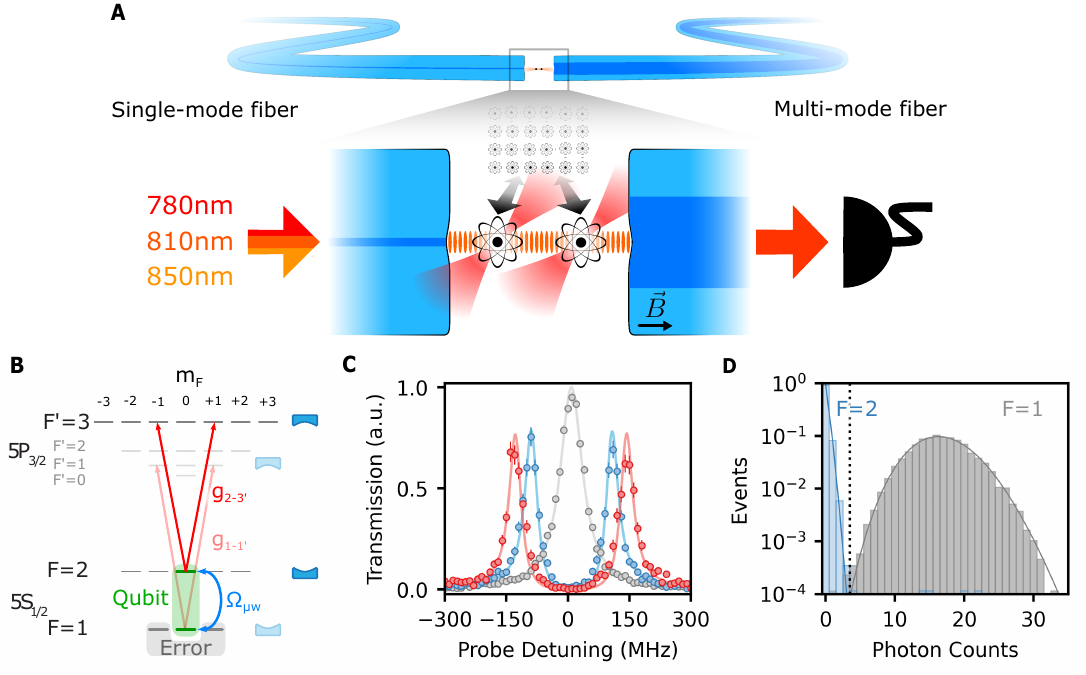}}
\caption{\textbf{Experimental setup and qubit readout.} \textbf{(A)}  Atoms in optical tweezers are loaded above a Fabry-Perot Fiber Cavity (FPFC) and transported into the cavity mode. The FPFC has a single-mode input port and a multi-mode output port, over which we collect transmitted light. The FPFC length is stabilized using an \SI{810}{\nano\meter} laser and the atoms are trapped in a \SI{850}{\nano\meter} mode of the cavity. The \SI{780}{\nano\meter} mode couples to the D2 line in Rubidium. The magnetic field is aligned along the cavity axis. \textbf{(B)} Atomic level diagram of Rubidium-87. Readout is performed with the cavity on resonance with the  $2\leftrightarrow3'$ transition, while for entanglement generation the cavity is tuned to the $1\leftrightarrow1'$ transition. \textbf{(C)} Resonant transmission spectra of a bare cavity (gray), one atom (blue), and two atoms (red) coupled to the cavity. The lines are a simultaneous fit yielding parameters g = 100.0(8) MHz and $\kappa$ = 65(1) MHz, resulting in a cooperativity of $C=101(2)$. \textbf{(D)} Histograms showing photon counts collected in \SI{10}{\micro\second} with an atom in the F=1 state (grey) and an atom in the F=2 state (blue), yielding a qubit state readout fidelity of $99.960^{+14}_{-24}\SI{}{\percent}$. All errorbars are SEM.}
\label{fig:fig1} 
\end{figure*}
 \paragraph*{Atom-cavity platform} We couple individually controlled Rubidium-87 atoms trapped in optical tweezers to a Fabry-Perot Fiber Cavity (FPFC). These micro-cavities exhibit low scattering loss and small radii of curvature, which allow for a small mode volume and high cooperativity~\cite{hunger2010fiber}. Our FPFC design features mirror diameters of \SI{125}{\micro\meter} and a cavity length of \SI{100}{\micro\meter}, providing ample optical access for single atoms in optical tweezers \cite{Supp}. We load atoms into optical tweezers from a magneto-optical trap (MOT) formed directly above the FPFC. The atoms are then transported approximately \SI{60}{\micro\meter} into the cavity mode and deposited into an optical lattice formed by an \SI{850}{\nano\meter} cavity mode (Fig.\refsub{fig:fig1}{A}). The atoms are positioned in the \SI{850}{\nano\meter} mode to maximize their overlap with the \SI{780}{\nano\meter} mode. We stabilize the FPFC using an \SI{810}{\nano\meter} laser referenced to an external ultra-low expansion (ULE) cavity \cite{Supp}, ensuring that the \SI{780}{\nano\meter} cavity mode remains locked to the $\ket{5S_{1/2}}\leftrightarrow\ket{5P_{3/2}}$ $D2$ transition of Rubidium-87. Additionally, the cavity resonance can be dynamically tuned across the entire hyperfine structure during an experimental sequence (Fig.\refsub{fig:fig1}{B}).
 
 The atom-cavity system is characterized by probing the transmission spectrum of the cavity through the single-mode input port using circularly polarized  ($\sigma^+$ ) light. By pumping the atoms to the stretched state $\ket{F = 2, m_F = 2}$ with the cavity resonant to the $F = 2\leftrightarrow F' = 3'$ transition, we realize an effective two-level system. The excitation spectrum of a single atom showcases a strong resolvable vacuum-Rabi splitting, showing the hybridization of the atomic and photonic excitations (Fig.~\refsub{fig:fig1}{C}). Fitting the transmission spectrum to an analytical model~\cite{Supp}, we extract an atom-photon coupling strength of $g=\SI{100.0(8)}{\mega\hertz}$ and cavity linewidth (FWHM) of $\kappa=\SI{65(1)}{\mega\hertz}$. This results in a single-atom cooperativity of $C=\frac{4g^2}{\kappa \Gamma} = 101(2)$, where $\Gamma = \SI{6}{\mega\hertz}$ is the natural linewidth of the excited state.

When probed on resonance, the difference in transmission between an atom coupled and no atom coupled to the cavity allows for fast non-destructive readout by thresholding transmitted photon counts. \cite{ gehr2010cavity, bochmann2010lossless, dorantes2017readout, deist2022mid} Through this measurement, atom presence can be detected with a fidelity of $99.988^{+7}_{-23}\SI{}{\percent}$ in \SI{10}{\micro\second} and of $99.950^{+24}_{-49}\SI{}{\percent}$ in \SI{2.5}{\micro\second}. Additionally, this readout is nearly lossless with a measured loss of $0.034^{+27}_{-6}\SI{}{\percent}$.
In all of our experiments, we post-select on atomic presence at the end of the experimental sequence~\cite{Supp}. Furthermore, this measurement can differentiate atoms in the qubit manifold, which we encode in the magnetic-field insensitive states: $\ket{0} = \ket{F=1,m_F= 0}$ and $\ket{1} = \ket{F=2,m_F= 0}$. During readout, the cavity is tuned to the $2\leftrightarrow 3'$ transition, so that the $|1\rangle$ state is coupled to the cavity, while the $|0\rangle$ state is not. Using the state dependent coupling we perform fast non-destructive readout of the qubit state with fidelity of $99.960^{+14}_{-24}\SI{}{\percent}$ in \SI{10}{\micro\second}, limited by off-resonant scattering from the \SI{850}{\nano\meter} trap and state preparation \cite{Supp}. Importantly, readout leaves the $\ket{0}$ state undisturbed while it quickly pumps the $\ket{1}$ state to the stretched $\ket{F=2,m_F=2}$ state. In principle, a Raman pulse can be used afterward to reinitialize the atom in $\ket{1}$, making this readout non-destructive to the qubit state.~\cite{Supp}. 

 \paragraph*{Photon-mediated entanglement through dark states}
 
We realize quantum entanglement  between two atoms coupled to a cavity using a photonic dark state. This state is an antisymmetric superposition where one atom is excited out of phase with the other, leading to destructive interference in the cavity mode, rendering it dark to the cavity. While such states have been studied previously \cite{neuzner2016interference, yan23superradiant}, their potential for quantum operations remains largely unexplored.

\begin{figure}[h!]
\centering
\includegraphics[width=3.5in]{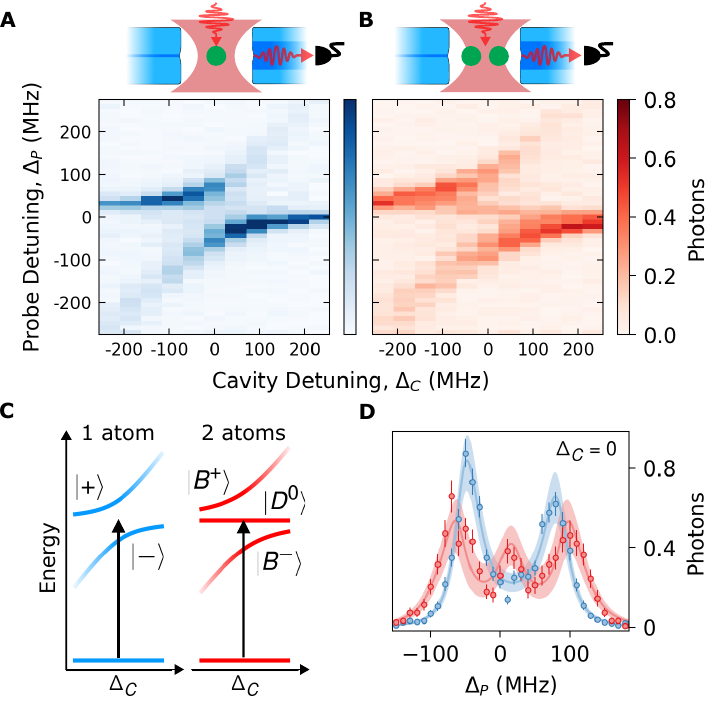}
\caption{\textbf{Photonic dark state spectroscopy.} \textbf{(A-B)} Side drive spectroscopy of one (A) and two (B) atoms coupled to the cavity as a function of cavity and probe laser detuning ($\Delta_C$ and $\Delta_P$, respectively). With two atoms coupled to the cavity, the spectrum reveals a feature on resonance with the atomic transition, corresponding to a photonic dark state. \textbf{(C)}  Energy level diagram of the first excitation manifold for one and two atoms coupled to a single cavity mode, which correspond to the states probed in the spectra plotted against the cavity detuning, $\Delta_c$~\cite{Supp}. \textbf{(D)} Probe spectroscopy at $\Delta_C = \Delta_A$ for one (blue circles) and two (red circles) atoms coupled to the cavity while probed with a side beam. Lines are simultaneous fits to a numerical model~\cite{Supp}. All errorbars are SEM.} 
\label{fig:fig2}
\end{figure}

Specifically, we probe the system by exciting both atoms from the side, rather than through the single-mode input port, and measure the same signal: the light transmitted through the multi-mode optical fiber. We vary both the cavity detuning $\Delta_C$ and the probe detuning $\Delta_P$. With a single atom coupled to the cavity, we directly observe the  same avoided-crossing seen before (Fig.~\refsub{fig:fig1}{C}). Moreover, the dressed states become more  atom-like (cavity-like) close to the atomic (cavity) resonance, resulting in stronger (weaker) coupling to the side beam drive (Fig.~\refsub{fig:fig2}{A}). With two atoms coupled to the cavity, similar behavior is observed but with an expected $\sqrt{2}$ enhancement of the splitting (Fig.~\refsub{fig:fig2}{B}). These two features correspond to the "bright states", given by $\ket{B_{\pm}} = \frac{1}{2}(\ket{eg, 0} + \ket{ge, 0})\pm\frac{1}{\sqrt{2}}\ket{gg, 1}$. In contrast to the spectrum recorded when the cavity mode is excited through the fiber, an additional third feature appears at the atomic resonance that we attribute to a photonic dark state, $\ket{D_0} = \frac{1}{\sqrt{2}}(\ket{eg, 0} - \ket{ge, 0})$ (Fig.~\refsub{fig:fig2}{C}), where $\ket{g}$ and $\ket{e}$ are the atomic ground and excited states of each atom and the number indicates the number of photons in the cavity. Due to the lack of a photonic component, this state can only be driven by exciting the atom and not the cavity. Specifically, for a global drive, a relative phase of $\pi$ between the drive on the two atoms is required. Due to thermal motion of the atoms, the relative phase between the two atoms and the drive beam changes from shot to shot, and we consistently observe the dark state as a bright feature. Furthermore, a strong drive allows for excitation of the cavity through the dark state. In the rest of our experiments, we individually drive single atoms, which maintains a coupling to both dark and bright states, while being insensitive to the relative phase between the atoms~\cite{Supp}.

To use this dark state to implement cavity-mediated entanglement between the two atoms, we note that  when the cavity is resonant with the bare atomic transition, the difference between the single- and two-atom spectra can be interpreted as a conditional resonance. The system can only be excited with a resonant beam if both atoms are coupled to the cavity (Fig.~\refsub{fig:fig2}{D}). By encoding our qubit in states that are coupled or uncoupled to the cavity, this conditional resonance becomes state dependent, enabling two-qubit quantum operations. However, the fidelity of these operations is limited by errors arising from photon leakage out of the cavity and scattering into free space. We can improve this fidelity by engineering these leakage and scattering events to leave the atom in a specific state, $\ket{err}$, outside the qubit manifold. Through sequences of readout, $\pi$-pulses, and local optical pumping, we can measure the qubit states $\ket{0}$ and $\ket{1}$, as well as $\ket{err}$ ~\cite{Supp}. By postselecting on cases where no error occurred, the fidelity will ultimately be limited by the presence of undetectable errors. \cite{borregaard2015heralded, ramette2024counter}

\vspace*{\fill}
\begin{figure}[h!]
\centering 
\includegraphics[width=3.5in]{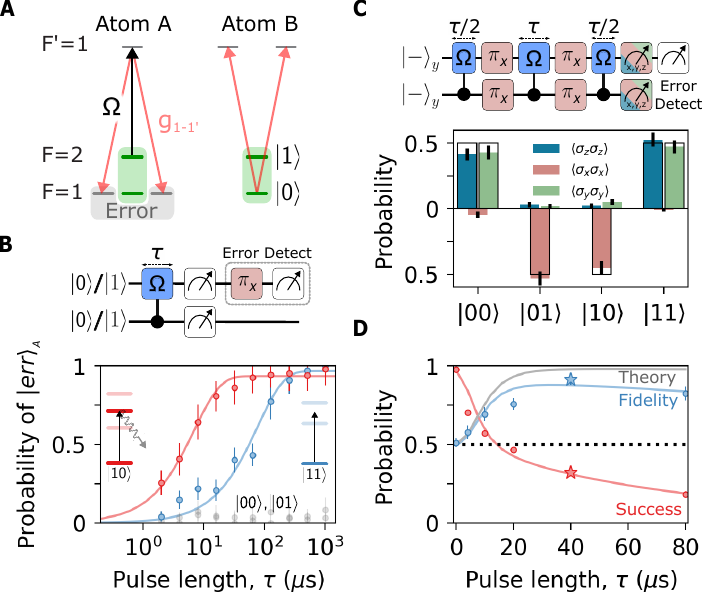}
\caption{\textbf{Bell state preparation with cavity carving.} \textbf{(A)} For entanglement generation, the cavity is tuned to the $1\leftrightarrow 1'$ transition and a single side drive on the $2\leftrightarrow 1'$ transition is applied to atom A only. \textbf{(B)} (top) Quantum circuit for preparing two-qubit states and measuring the $\ket{err}_A$ population. (bottom) Decay rates into the error states after preparing each of the four two-qubit states. The resonant coupling to the dark state $\ket{D_0}$ results in much faster decay of $\ket{10}$ compared to the off-resonant coupling in $\ket{11}$. $\ket{00}$ and $\ket{01}$ do not couple to the driving beam. \textbf{(C)} (top) Quantum circuit of a Carr-Purcell cavity carving scheme which prepares a $\ket{\Phi^-}$ Bell state. (bottom) Measured correlations in the ZZ (blue), XX (red), and YY (green) bases verify that $\ket{\Phi^-}$ is prepared with a Bell-state fidelity of $\mathcal{F}=\SI{91(4)}{\percent}$ and a success probability of \SI{32(1)}{\percent}. Direction of histograms denote the expected sign of parity for $\ket{\Phi^-}$ in each bases. \textbf{(D)} Bell-state fidelity (blue circles) and success probability (red circles) as a function of carving pulse length $\tau$. Stars mark the data presented in (C). Lines are a fit to a semi-classical numerical model, and the gray line is the expected Bell-state fidelity without SPAM errors~\cite{Supp}. Dotted line denotes the fidelity limit for verifiable entanglement. All errorbars are SEM.} 
\label{fig:fig3}
\end{figure}

For entanglement experiments, we use the excited state $\ket{e}=\ket{F'=1, m_F=0}$ which has branching ratios that result in decay predominantly to the $\ket{err}=\ket{F=1, m_F=\pm1}$ states, while selection rules forbid decay to the $\ket{0}$ state (Fig.~\refsub{fig:fig3}{A}). We perform a local drive on one of the atoms, which we call atom A, resonant with the $2\leftrightarrow 1'$ transition. This drive couples $|1\rangle_A$ to $|e\rangle_A$. We shift the cavity resonance to the $1 \leftrightarrow 1'$ transition, so that it couples $\ket{err}_A \leftrightarrow \ket{e}_A$ and $\ket{0}_B \leftrightarrow \ket{e'}_B$ where $\ket{e'} = \ket{F' = 1, M_F = \pm 1 }$ state. As a result, the two-qubit state $\ket{1}_A\ket{1}_B\equiv\ket{11}$ acts as a single atom coupled to the cavity and experiences a suppressed excitation. By contrast, the $\ket{10}$ state behaves as two atoms coupled to the cavity and experiences a resonant coupling to the photonic dark state, $\ket{D_0}=1/\sqrt{2} \left(\ket{err}_A\ket{e'}_B-\ket{e}_A\ket{0}_B \right)$~\cite{Supp}. This establishes a qubit state-dependent resonance, necessary to perform quantum operations between the two qubits.

Our first protocol relies on the state-dependent decay of the two-qubit states. Under a weak drive, we observe that the $\ket{11}$ and $\ket{10}$ states both decay into the $\ket{err}_A$ state with different rates. The resonant coupling to the dark state makes the $\ket{10}$ state rapidly decay, while the $\ket{11}$ experiences a suppressed decay (Fig.~\refsub{fig:fig3}{B}). This allows us to effectively "carve out" the $\ket{10}$ state from the wavefunction. \cite{welte2018gate, dhordjevic2021entanglement, ramette2024counter} To prepare an entangled state with this mechanism, we first prepare an equal superposition of all two-qubit states, followed by the application of a drive pulse which carves out the $\ket{10}$ component. A global $\pi$-pulse flips the two-qubit state and a second carving pulse once again carves out the $\ket{10}$ component. This sequence prepares a statistical mixture of error states and a Bell state: $\rho=\ket{\Phi^-}\bra{\Phi^-}+\ket{err}_A\bra{err}$. By post-selecting on the atoms not being in the error state, we are left with the maximally entangled state, $\ket{\Phi^-} = \frac{1}{\sqrt{2}} \left(\ket{00} - \ket{11} \right)$. In principle, the minimal application of our scheme would be the spin-echo sequence described above: $\frac{\pi}{2}\text{-}\tau\text{-}\pi\text{-}\tau\text{-}\frac{\pi}{2}$. In practice, we found that the spin-echo scheme was limited due to a coherent linear phase accumulation in the $\ket{11}$ state during the carving pulses that we attribute to laser phase noise~\cite{Supp}. To cancel this linear phase, we implement a Carr-Purcell decoupling sequence of the form $\frac{\pi}{2}\text{-}\frac{\tau}{2}\text{-}\pi\text{-}\tau\text{-}\pi\text{-}\frac{\tau}{2}\text{-}\frac{\pi}{2}$. Fig.~\refsub{fig:fig3}{C} shows the measured correlations in the $ZZ$, $XX$, and $YY$ basis, resulting in a Bell-state fidelity of $\mathcal{F}=\SI{91(4)}{\percent}$ and a success probability of \SI{32(1)}{\percent}. 

The theoretical limit on fidelity is set by undetectable errors, i.e. when a scattering event leaves the atoms in the qubit manifold instead of in the error state. The dominant source of undetectable errors arise from scattering from the dark state, which can decay through atom A or through atom B with equal probability. Decay through atom B leaves atom A in the state $\ket{err}_A$, so that this decay is fully detectable. Decay through atom A result in $\SI{17}{\percent}$ undetectable errors for a single atom given by Clebsch-Gordon coefficients and due to our current readout sequence, which cannot distinguish the F=2 Zeeman sublevels. This leads to total dark state scattering errors being $\SI{91.5}{\percent}$ detectable.
With this, we find a maximum theoretical fidelity of $\sim\SI{96.3}{\percent}$ for our carving protocol. As predicted by our model, Fig.~\refsub{fig:fig3}{D} shows that the entanglement fidelity increases exponentially, but is still limited by unwanted scattering which we attribute to level-mixing of the excited states from our traps, and imperfect polarization~\cite{Supp}.

\paragraph*{Quantum operations with integrated error detection}
\begin{figure}
\centering
\includegraphics[width=2.2in]{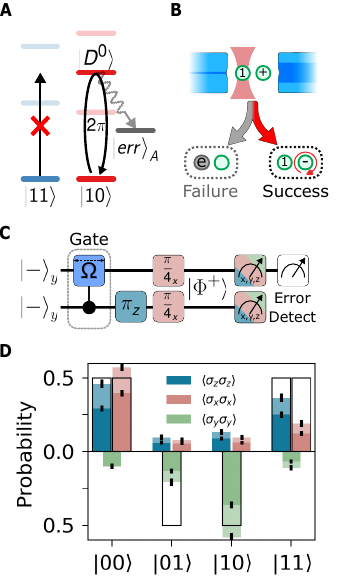}
\caption{\textbf{Deterministic quantum gate with error detection.} \textbf{(A)} Conditional blockade mechanism. A resonant $2\pi$ Rabi oscillation can be realized only between the $\ket{10}$ state and the dark state $\ket{D_0}$, causing a phase of $\pi$ on state $\ket{10}$, while the excitation from the $\ket{11}$ state is blockaded. Unwanted atomic scattering results in mainly detectable $\ket{err}_A$ states.  \textbf{(B)} Quantum gate flow chart showing the two possible outcomes: "Success" is a conditional phase on atom B and "Failure" is a detectable scattering event \textbf{(C)} Quantum circuit of the gate: Following the gate, we apply a local $Z_{\pi}$ pulse on atom B and a global microwave $X_{\pi/4}$ pulse to generate the $\ket{\Phi^+}$ Bell state. \textbf{(D)} Measured correlations in the ZZ (blue), XX (red), and YY (green) bases verify that $\ket{\Phi^+}$ is prepared. (dark) Without error detection, we measure a deterministic Bell-state fidelity of $\mathcal{F}=\SI{52.5(1.8)}{\percent}$. (light) Error detection improves the fidelity to $\mathcal{F}'=\SI{76(2)}{\percent}$ with a success probability of \SI{69(1)}{\percent}. Direction of histograms denote the expected sign of parity for $\ket{\Phi^+}$ in each bases. All errorbars are SEM.}
\label{fig:fig4}
\end{figure}
The qubit state dependent resonance can also be used to perform a quantum gate between two atoms. By increasing the strength of the laser, we can drive a coherent optical Rabi oscillation between the $\ket{10}$ state and the photonic dark state $\ket{D_0}$, while the $\ket{11}$ state is blockaded from excitation and both the $\ket{00}$ and $\ket{01}$ states are unaffected by the drive. By performing a full $2\pi$ rotation, the $\ket{10}$ will acquire a relative $\pi$ phase compared to the other three two-qubit states, realizing a controlled phase gate (Fig.~\refsub{fig:fig4}{A}). This scheme is limited by scattering, however most of these events result in detectable error states, $\ket{err}_A$, allowing us to increase the fidelity of the measurement through post-selection (Fig.~\refsub{fig:fig4}{B}). 

To characterize the performance of the gate, we prepare an entangled state using the quantum circuit shown in Fig.\refsub{fig:fig4}{C}. For a perfect gate, this circuit would prepare the Bell state $\ket{\Phi^+} = \frac{1}{\sqrt{2}} \left(\ket{00} + \ket{11} \right)$. Similar to the carving case, we determine the Bell-state fidelity by measuring correlations along all three bases (Fig.\refsub{fig:fig4}{D}). Notably, without any error detection or post-selection, the measured Bell-state fidelity is $\mathcal{F}=\SI{52.5(1.8)}{\percent}$~\cite{Supp}. By applying error detection and post-selection, we improve the Bell-state fidelity to $\mathcal{F}'=\SI{76(2)}{\percent}$, with a success probability of \SI{69(1)}{\percent}. This measurement aligns with the theoretical maximum corrected fidelity of \SI{78}{\percent} for our cooperativity. The limitation in fidelity is primarily due to the use of atom A as both a qubit and an ancilla to herald the success of the gate, as well as the fact that the states used in the current scheme do not have fully detectable errors. In addition, the protocol suffers from scattering and a light shift from the F=3 states that results in a further reduction in fidelity. These limitations can be addressed in future implementations by including an additional ancillary atom and by more careful selection of atomic states~\cite{borregaard2015heralded,ramette2024counter,Supp}.  Such an ancilla could be individually addressed and read out, allowing for the detection of gate errors without affecting the data qubits. 

\paragraph*{Outlook} Our experiments demonstrate  a versatile platform that offers exciting opportunities by combining neutral atom arrays with a high-finesse optical cavity. The demonstrated entanglement schemes offer significant improvements over many existing quantum networking protocols. For instance, the probabilistic gate performed using photon detection, as in \cite{langenfeld2021quantum}, can be replaced with cavity carving that includes error detection of the atomic state, potentially increasing the rate of quantum information transmission by at least  an order of magnitude. Another extension  would be to increase the number of atoms in the cavity, introducing an auxiliary atom to serve as an ancilla for the gate, thereby enhancing gate fidelity \cite{borregaard2015heralded}. Moreover, incorporating more qubits would allow for the execution of multi-qubit gates \cite{jandura2023non}. Higher fidelity cavity-mediated gates could unlock new possibilities in quantum networking by utilizing many entangled matter qubits, potentially generating many-photon graph states for error-corrected quantum networking protocols \cite{borregaard20oneway, thomas2024fusion}. Further improvements can be achieved by fabricating higher-quality mirrors to increase cooperativity \cite{watcher19micromirror, jin2022micro}. Combining our platform with Rydberg gate operations in atom arrays would enable fast, high-fidelity, non-destructive mid-circuit readout for error-correction protocols and facilitate the entanglement of spatially separated quantum processors for distributed quantum computing, thereby increasing the available number of qubits \cite{huie2021multiplexed, young2022architecture, sinclair2024fault, li24highrate}. Finally, our error-biased mechanism can be extended to more general cavity-mediated interactions, such as long-range spin-spin Hamiltonians and spin squeezing \cite{leroux2010implementation, cox2016deterministic, hung2016quantum, periwal2021programmable, norcia2018cavity}.

\textbf{Acknowledgments:} We thank Polnop Samutpraphoot, Tamara Dordevic, and Paloma Ocola for assisting in the early stages of building this experiment, Lilian Childress,  Jiaxing Ma, Thomas Clark, and Sylvain Schwartz for helpful discussion on fiber cavity mounting and alignment,  Michael Förg and Thomas Hummer for useful discussions about fiber cavities,  Gefen Baranes and Elias Trapp for insightful discussions, and  Andrei Ruskuc for critical reading of the manuscript.
\textbf{Funding:}
This work was supported by the DOE QSA Center (DE-AC02-05CH11231), the National Science Foundation (grant number PHY-2012023), the Center for Ultracold Atoms (an NSF Physics Frontiers Center), ARO MURI (W911NF2010082), and DARPA ONISQ (grant number W911NF2010021). The device was fabricated at the Harvard CNS (NSF ECCS-1541959).
\textbf{Author contributions:} 
All work was supervised by V.V. and M.D.L. All authors 
discussed the results and contributed to the manuscript. 
\textbf{Competing interests:} 
V.V. and M.D.L. are co-founders and shareholders of QuEra Computing.
\textbf{Data and materials availability:} All data needed to evaluate the conclusions in the paper are present in the paper and the supplementary materials.
 \nocite{ Rudelis2023MirrorDegradation, abraham1998teflon, tiecke2014switch, qutip, Reiter2012} 
 
\bibliography{bib.bib}
\bibliographystyle{Science}

\clearpage

\renewcommand{\thefigure}{S\arabic{figure}}
\renewcommand{\thetable}{S\arabic{table}}
\setcounter{equation}{0}
\setcounter{figure}{0}
\setcounter{table}{0}

\begin{center}
{\LARGE Supplementary Materials}
\end{center}
{\let\newpage\relax\maketitle}

\tableofcontents
\section{Fiber Cavity Mounting and Parameters} 

\begin{figure}
\centering
\includegraphics[width=.45 \textwidth]{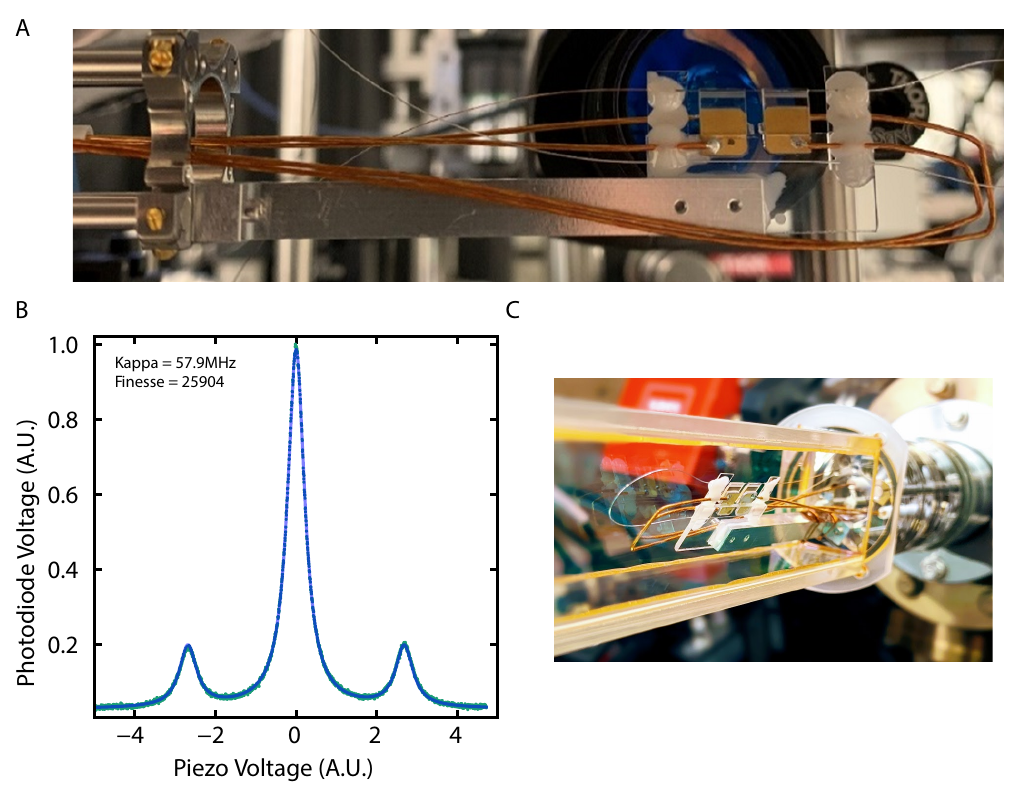}
\caption{{\bfseries FPFC Mounting and Characterization.} (A) Photo of the mounting structure for the Fabry-Perot Fiber Cavity. (B) Cavity transmission {\it vs.} piezo voltage (blue circles) when probed with resonant light using \SI{300}{\mega\hertz} sidebands. A fit to 3 Lorentzians (blue line) reveals the cavity linewidth $\kappa = \SI{57.9}{\mega\hertz}$ and finesse $\mathcal{F}_{cav}=25904$ assuming a \SI{100}{\micro\meter} length. (C) The mounting structure for the Fabry-Perot Fiber Cavity in vacuum}
\label{fig:cavitychar} 
\end{figure}

Our fiber cavities are purchased from the company, Qlibri, with a coating with a reflectivity $> 99.993\%$ centered around \SI{780}{\nano\meter}. A protective layer of SiO$_2$ is added as a final layer to the coating to prevent degradation in vacuum~\cite{Rudelis2023MirrorDegradation}. We work with a single-mode (SM) fiber (IVG Cu800) with a radius of curvature of (ROC) of \SI{165}{\micro\meter} for the input mirror. For the output mirror, we use a multi-mode (MM) fiber (Art Photonics MM50) with an ROC of \SI{145}{\micro\meter}. Each fiber has a diameter of \SI{125}{\micro\meter}. Our tweezers are formed with a 0.5 NA objective (Mitutoyo G Plan Apo 50X Objective). To allow for enough optical access for the optical tweezers, we choose a cavity length of \SI{100}{\micro\meter}. The fiber cavities are mounted on a monolithic glass V-groove structure glued onto two shear piezo (CSAP02), which are glued on top of a glass slide. The fiber cavities are mounted into the V-grooves, then low expansion glue (Optocast 3410 Gen2-40K) is applied to the fibers. This is then mounted into an aluminum structure that secures the glass slide, as shown in the photos of Fig.~\refsub{fig:cavitychar}{A}. 

Both fibers are guided out of the vacuum chamber through a fiber feed-through. The fiber feed-through is composed of a two-hole Teflon piece tightened into a Swagelok \cite{abraham1998teflon}. Upon placement in the vacuum chamber, the finesse of the cavity exceeded \SI{40000}{}. During the bake-out, a sudden change in temperature caused the cavity to become misaligned. This resulted in a misalignment of the cavity that reduced the finesse to $\sim\SI{25000}{}$. Additionally, the misalignment resulted in a reduction of coupling between the mode out of the fiber and the mode of the cavity. 

Finesse is measured by producing \SI{300}{\mega\hertz} sidebands on the cavity mode using a fiber electro-optical modulator (EOM). These sidebands can then be used to measure the cavity linewidth by providing a way to transform piezo voltage into megahertz, assuming linear response. With this measurement and a separate measurement of the length of the cavity we can calculate a finesse of $F = 25,904$ as shown in Fig.~\refsub{fig:cavitychar}{B}.

\begin{figure*}
\centering
\noindent\makebox[\textwidth]{\includegraphics[width=\textwidth]{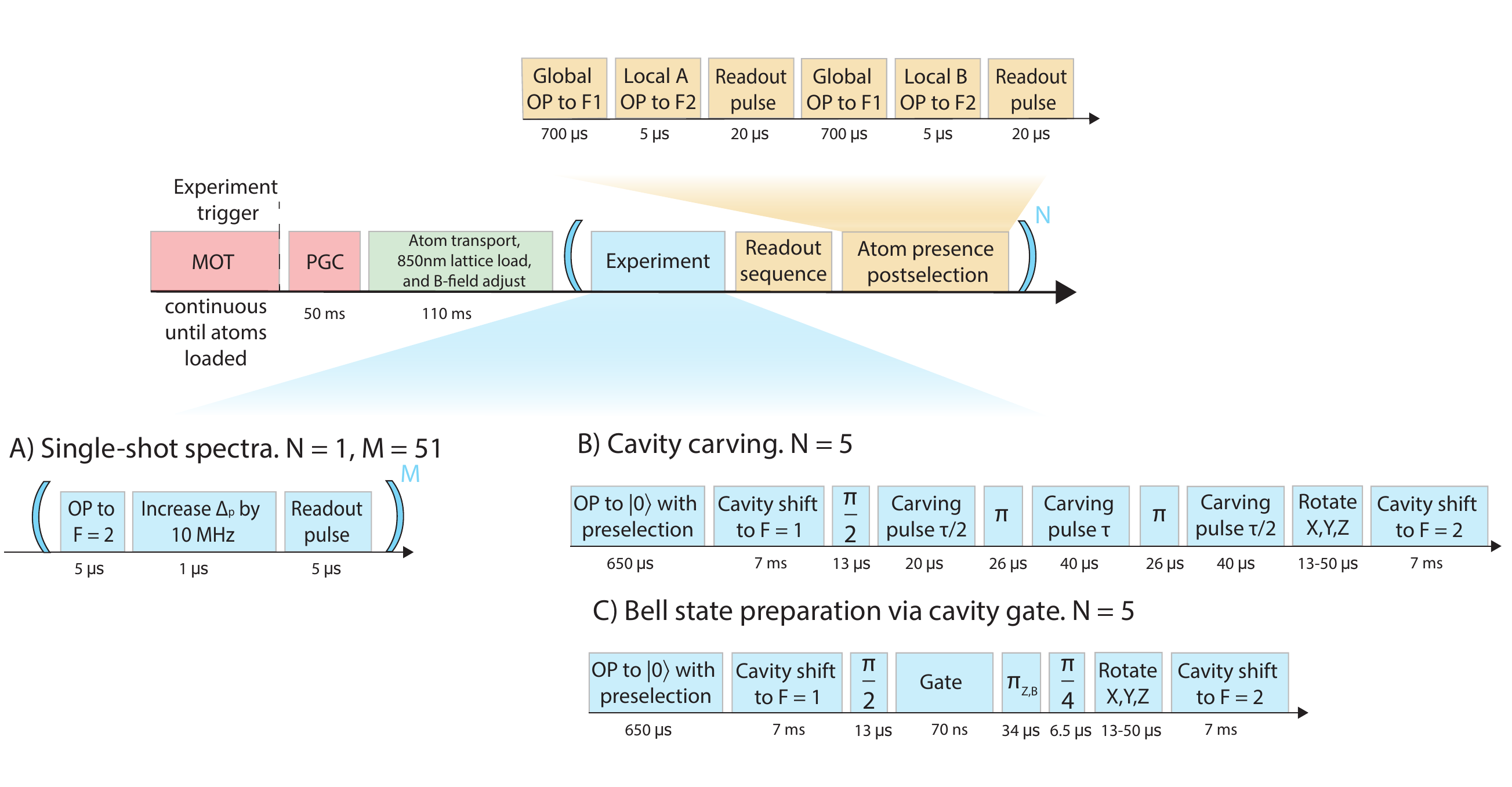}}
\caption{Experimental Sequence for: A) the experiments presented in Figs. 1-2 in the main text; B) cavity carving and the data in Fig. 3 in the main text; and C) Bell state preparation using an error-detected gate and the data in Fig. 4 in the main text. }
\label{fig:experimentseqs} 
\end{figure*}

\section{Experimental Methods}

\subsection{MOT and Tweezer Loading}

Atoms are loaded in optical tweezers from a Magneto Optical Trap (MOT) which is formed right above the FPFC. Fluorescence from atoms loaded in the tweezers is collected on single-photon avalanche detectors (SPADs) co-aligned with the tweezers. However, scattering of the MOT beams from the cavity and its mounts prevents the MOT from forming within the cavity and also overwhelms the SPADs. To overcome these problems, we clip the horizontal MOT beams using razor blades placed in MOT beam paths. We fine-tune their position to optimize atom loading into the tweezers. In addition, to reduce background counts from the light scattering by the MOT beams off the FPFC, we include an additional imaging beam that propagates along the MOT beam perpendicular to the FPFC. This beam is focused through the cavity such that there is very little scattering. This beam is then pulsed out of phase with the MOT and the tweezers at \SI{2}{\mega\hertz}. The experimental sequence for each of the experiments presented in the main text is shown in Fig.~\refsub{fig:experimentseqs}. The sequence begins with polarization gradient cooling (PGC) to reduce the temperature of the atoms. The atoms are then transported into the cavity mode.

\subsection{Atom Positioning in The Fiber Cavity}

\begin{figure}[h!]
\centering
\includegraphics[scale=0.35]{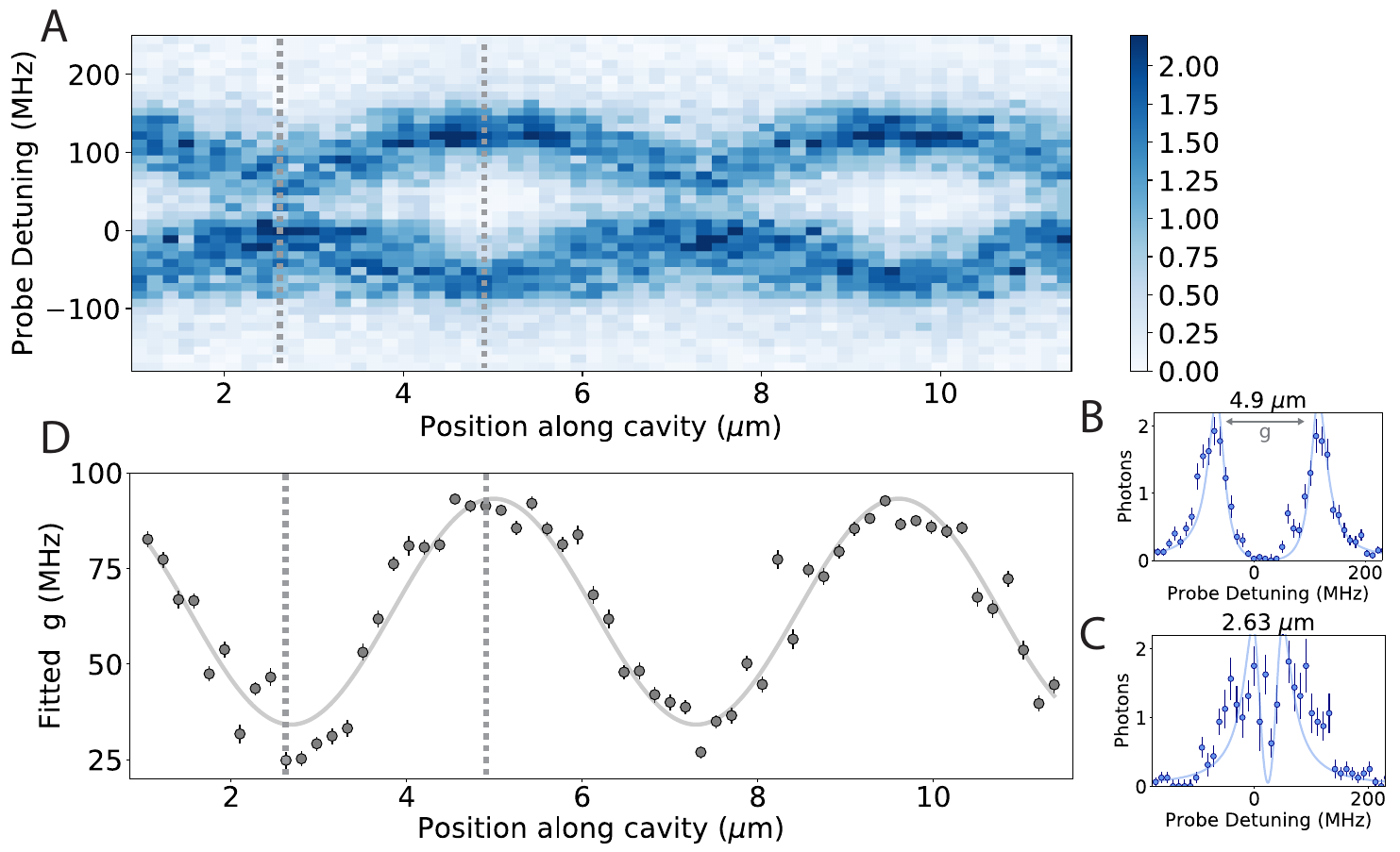}
\caption{Atom positioning in the cavity mode. \textbf{A} Single-shot spectra (averaged) of single atoms placed at different positions along the cavity. \textbf{B} and \textbf{C} Two such spectra at positions of minimum (2.64 um) and maximum (4.95 um) coupling. The solid lines are fits of the form $T(\delta)$ in \eqref{eqn:probespectra}. \textbf{D} Fitted atom-cavity coupling strength $g$ as a function of position along the cavity. The solid line is a fit of the form $g(x) = a \cos(2 \pi (x-x_0)/P) + c$. Note that we use the fitted period $P$ to calibrate the distance along the cavity.   } 
\label{fig:scanx} 
\end{figure}

In order to ensure that the atoms are always coupled to the cavity mode, we use an additional optical lattice formed by a 850nm laser which is resonant with a cavity mode. The atoms are transported by the tweezers using galvo mirrors and are loaded into the 850nm lattice. The 850nm and the 780nm modes are incommensurate and form a beat note with a period of \SI{4.7}{\micro\meter} with regions where their maxima overlap. We position each of the tweezers in one such region. Fig.~\refsub{fig:scanx}, shows transmission spectra when a single atom is placed at different positions along the lattice. The regions of maximum splitting between the hybridized atom-photon modes correspond to maximum coupling $g$ and we place the atoms there. This allows us to place around 20 atoms in the cavity, and more if we accommodate more than one atom in each beat note.

\subsection{State Preparation}

Atoms are prepared in the $\ket{0}=\ket{F=1,m_F=0}$ by applying a linearly polarized drive on the $\ket{5S_{1/2},F=1}\leftrightarrow \ket{5P_{1/2},F'=1}$ D1 transition while simultaneously applying a drive on the $\ket{5S_{1/2},F=2}\leftrightarrow\ket{5P_{3/2},F'=1}$ D2 transition with mixed polarization. With these drives, the $\ket{F=1,m_F=0}$ is a dark state, and the population is pumped there in $\SI{15}{\micro\second}$. Imperfect polarization limits this state preparation. To improve this scheme, we extend the duration of the D1 beam by $\SI{3}{\micro\second}$, pumping the residual population from the $\ket{F=1,m_F=\pm1}$ states in to the $F = 2$ manifold. We then apply a readout pulse, resonant with the $2 \rightarrow 3'$ transition and postselect and discard the data when an atom in $F = 2$ is detected. This allows us to preselect the cases when optical pumping (OP) was successful before we begin the experiment. Note that this readout pulse does not affect the $\ket{0}$ state. The final state preparation fidelity is $>\SI{99}{\percent}$.

\subsection{Cavity and Laser Locking and Jumping}

\begin{figure}[h!]
\centering
\includegraphics[width = 3.5in]{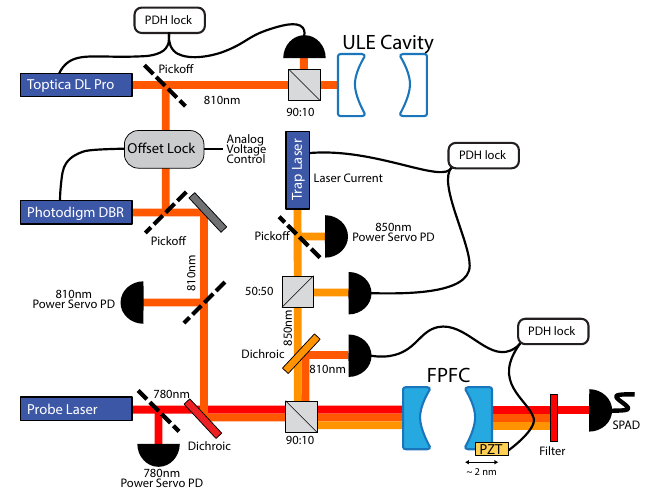}
\caption{{\bfseries FPFC laser system.} Diagram of the lasers coupled to the FPFC and the different locking systems for the cavity and each laser. Waveplates and fiber couplings have been ommitted.} 
\label{fig:cavityschematic} 
\end{figure}

Fig.~\refsub{fig:cavityschematic}~ shows a diagram of the cavity locking system. We stabilize the cavity length using a shear piezo and a Pound Drever Hall (PDH) lock, which uses the cavity reflection signal from an \SI{810}{\nano\meter} DBR laser (Photodigm 808DBRH-TOSA). The frequency of the DBR laser is locked and controlled using an offset phase lock (Vescent D2-135) to reference it with a \SI{810}{\nano\meter} ECDL laser (Toptica DL100 Pro ECDL). The ECDL laser, is locked to an ultra-low expansion (ULE) cavity using a PDH lock to act as a stable reference. This system allows us to dynamically tune the cavity length and resonance. Using an analog channel and tunable voltage controlled oscillators (VCOs), we can dynamically tune the relative frequency between the tunable DBR laser and the stable ECDL laser. Taking advantage of the large current controlled mode-hop free range of the DBR laser, we achieve a tunability of up to $\sim\SI{10}{\giga\hertz}$, which correspond to $\sim\SI{2}{\nano\meter}$ changes in the length of the cavity. By choosing correctly the reference line of the ULE cavity, we can match this tunable range to the D2 line of $^{87}$Rb when locking the FPFC to the DBR laser. This system allows us to dynamically tune the cavity resonance through the whole hyperfine splitting of $^{87}$Rb, which is \SI{6.837}{\giga\hertz}. 

Thanks to the small size of the FPFC, we can do this relatively quickly in $\sim\SI{7}{\milli\second}$, limited only by resonant frequencies related to the lever-arm mechanical modes of the dangling fibers. After learning how to load atoms directly from a MOT close to these types of micro-cavities, we plan to increase the bandwidth of a next iteration by reducing the lever-arm length in the design. 

An additional $\SI{850}{\nano\meter}$ DBR laser is used as a trap for the atoms, and it's continuously locked to the cavity with a PDH lock on the laser diode current. The circulating power of the \SI{850}{\nano\meter} is $\sim\SI{20}{\milli\watt}$ resulting in $\sim\SI{24}{\mega\hertz}$ lightshift on the $2\leftrightarrow3'$ transition and similar of magnitude trap depth. The circulating power of the \SI{810}{\nano\meter} lock laser is $<\SI{0.5}{\milli\watt}$ in comparison and has negligible trapping or lightshift effects. 

\section{Readout}
\subsection{Detector System and Characterization}
Readout is performed by measuring the difference in transmission with and without an atom coupled to the cavity. We collect the transmitted photons through the Multi-mode (MM) port of the FPFC. After filtering of the trap light, \SI{780}{\nano\meter} light is then split on a 50:50 MM fiber beamsplitter and sent to two SPCM-AQRH-14-FC single photon avalanche detectors (SPADs) which are counted using a Swabian Time Tagger Ultra. We use two SPADs to effectively reduce the detector dead-time, increasing the bandwidth, and allowing for faster readout. We measure the effective detector dead-time by fitting the measured count-rate $C_M$ versus the expected probe count-rate $C_P$ using the model:
\begin{align}
    C_M = \left(C_P+C_D\right)e^{-t_D\left(C_P+C_D\right)\tau_0}
\label{eqn:deadtime}
\end{align}

Where we find that the detector dead time to be $t_D=\SI{17(1)}{\nano\second}$, $C_D$ is the dark count-rate, and $\tau_0=\SI{50}{\micro\second}$ is the integrated time in the measurement (Fig.~\refsub{fig:readoutchar}{A}). The dead time limits the linearity of the detector to count rates above $\sim\SI{10}{\mega\cps}$. We find that the atom starts to be saturated well below this limit, even accounting for collection efficiency. Therefore, the readout is limited by the power with which we can drive the system before the atom is saturated, and we leave the weak-probe regime.

We directly measured the dark count-rate to be $\SI{3.14(6)}{\milli\cps}$, finding it to be much smaller than even the coupled atom count rates. As such, the dark count-rate can be ignored from the model.

\subsection{Readout Fidelity}
We perform a readout measurement by probing the cavity with a \SI{780}{\nano\meter} laser pulse that is $\sigma^+$ circularly polarized resonant with the $2\leftrightarrow3'$ cycling transition and the cavity. The polarization helps increase the readout fidelity, as it quickly pumps the atoms into the stretched $\ket{2,2}$. For example, when reading out the clock state $\ket{2,0}$, due to the degeneracy of the cavity, we quickly pump into the $\ket{2,2}$ stretched state at a rate set by $\kappa$ scattering a single photon. This is the result of coupling to a state given by  $g_{\sigma^+} \ket{2,0} \ket{1}_{\sigma^-} - g_{\sigma^-} \ket{2,2} \ket{1}_{\sigma^+}$. The lambda system that produces this state is depicted in \refsub{fig:readoutchar}{B}. This state decays at a rate set by the cavity, making this pumping more rapid than one would expect from the atomic scattering rate $\gamma$. Additionally, this pumping quickly places us into the cycling transition without any atomic excitation, allowing us to take advantage of the cycling transition in the system for the highest readout fidelity.

When counting photons on transmission, there are two possible rates we can measure. When the cavity is empty, or the atoms inside are in an uncoupled state ($F=1$ ground-state manifold), we observe a high (H) count rate $C_M$, shown in Fig.~\refsub{fig:readoutchar}{A}. When there are atoms inside the cavity in a coupled state ($F=2$ ground-state manifold), we observe a low (L) count rate $C_{Atom}$ as shown in Fig.~\refsub{fig:readoutchar}{C}. $C_{Atom}$ is about 2 orders of magnitude lower than $C_M$ as the coupled atom effectively detunes the excited atom-cavity manifold and blockades the probe light. Both rates depend on the input probe power, which is directly proportional to the implied probe count-rate $C_P$. In the case of the coupled atom, we commonly work at a probe rate of $C_P\approx \SI{1.75}{\mega\cps}$ which sits right at the onset of where high counts start being affected by the detector dead-time and the low counts leave the linear weak-probe regime. 

\subsubsection{Readout Characterization}
In practice, there are two different types of readout pulses we perform, which we call atom- and state-readout. The only difference between the two is whether we are concurrently optically pumping the atoms into the coupled state manifold. In this way, atom-readout measures presence by always having the atoms in the coupled state, and we use mostly for post-selection on atom survival. We characterize our readout fidelity by measuring a train of subsequent atom- and state-readout pulses initializing an atom in either coupled or uncoupled states using optical pumping, finalizing on an atom-readout pulse to post-select on atom presence. Fig.~\refsub{fig:readoutchar}{D} shows a 2D histogram of the photon counts of subsequent atom- and state- readout pulses for atoms in the coupled and uncoupled states, as well as an empty cavity for comparison. To get good statistics at the high fidelities we are measuring, we have to accrue thousands of data points. The data presented is the same as for Fig.~1D of the main text, it corresponds to $C_P\approx \SI{1.75}{\mega\cps}$ and an integration time of \SI{10}{\micro\second}, which gives us the best state-readout fidelity of $99.960^{+14}_{-24}\SI{}{\percent}$ presented in the main text and corresponds to the usual probe power used throughout our experiments.

To measure readout fidelity, we need to understand the nature of photon counts, which can be described by the Poissonian distribution $P(\mu,k) = \frac{\mu^k e^{-\mu}}{k!}$. The curves on the count histograms of Fig. 1D of the main text are fits to this distribution. The fits give us two possible distributions: Low counts for atoms in a coupled stated with $\mu_L=\SI{0.09(2)}{\photons}$ and High counts for atoms in an uncoupled state with $\mu_H = \SI{16.600(1)}{\photons}$. From these, we can calculate an optimal threshold by equating both distributions and finding an analytical form of $k_{T} = \left\lfloor \frac{\mu_H - \mu_L}{\log{\mu_H} - \log{\mu_L}} \right\rfloor = \SI{3}{\photons}$, where an atom will be labelled (L) as coupled if the measured counts are $\leq k_{T}$, or (H) and uncoupled otherwise. As such, we can theoretically estimate the expected False positive probability to be $P_{FP} = 1 - \sum_{k=0}^{k_T}P(\mu_L,k) = 2.7\times10^{-6}$ and the  expected False negative probability to be $P_{FN} =  \sum_{k=0}^{k_T}P(\mu_H,k) = 57\times10^{-6}$. Experimentally, we calculate the False positive and negative rates to be $P_{FP} = 4.6^{+3.6}_{-2.2}\times10^{-4}$ and $P_{FN} = 3.5^{+3.3}_{-1.9}\times10^{-4}$. For simplicity, we quote a single Infidelity probability $P_{Inf} = (P_{FP} + P_{FN})/2$ to be the average of the two, given an atom can only be in one of two states every readout pulse. The quoted readout fidelities are then $(1-P_{Inf})\times\SI{100}{\percent}$. The estimated error for these measurements is the Clopper-Pearson interval.

\subsubsection{Optimal integration time and Limitations}
\begin{figure*}
\centering
\noindent\makebox[\textwidth]{\includegraphics[width=6in]{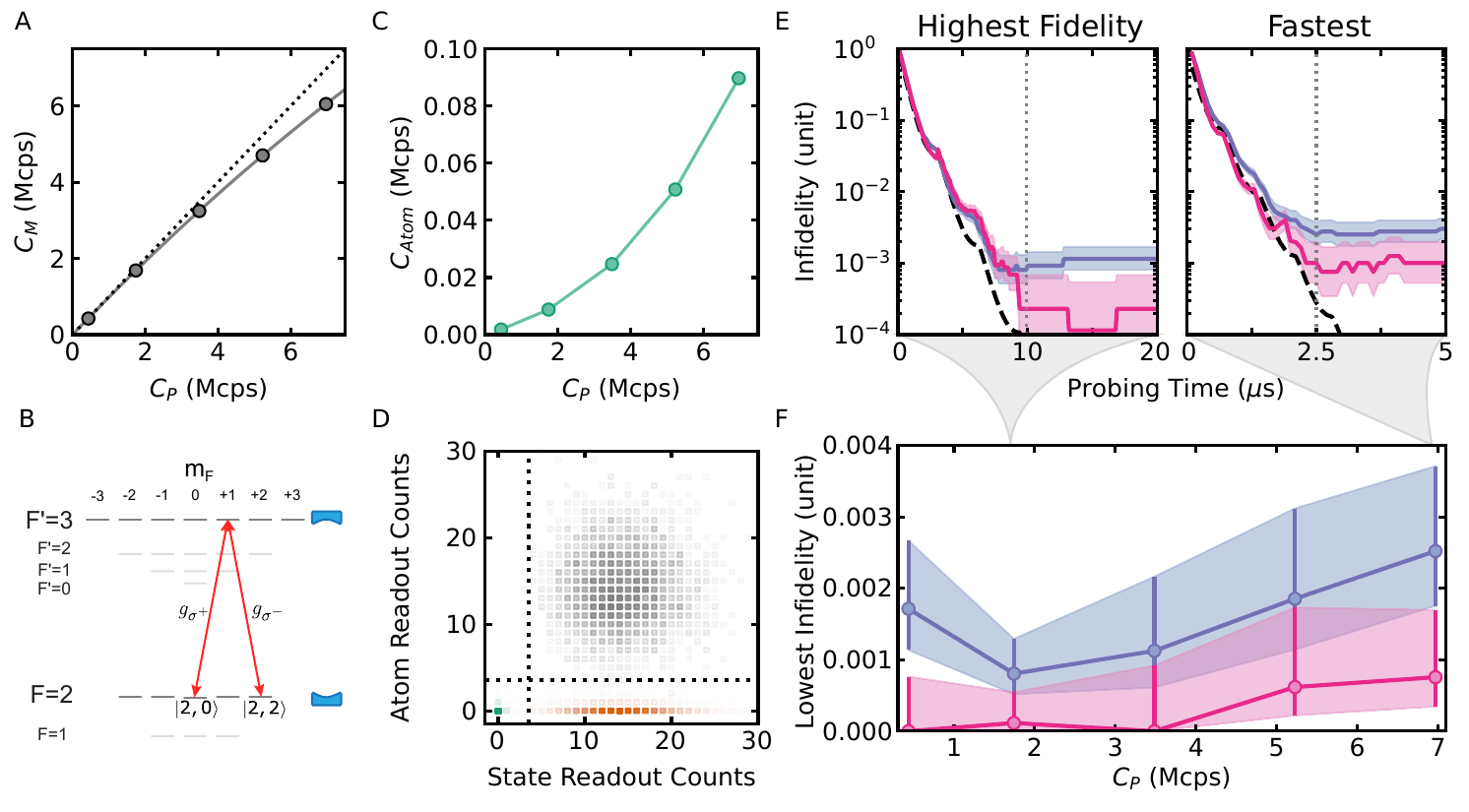}}
\caption{{\bfseries Readout Characterization.} (A) Measured count-rate $C_M$ {\it vs.} probe count-rate $C_P$ (gray circles) of our detection system with a fit (gray line) to model in Eqn.~\ref{eqn:deadtime} used to extract detector dead time. (B) Degenerate cavity allows for fast pumping into stretched state for enhanced coupling during readout. (C) Photon count-rate when with coupled atom $C_{Atom}$ {\it vs.} probe count-rate $C_P$ showing the breakdown of the linear weak-probe regime. (D) 2D histogram of readout pulse characterization, showing the counts of subsequent state- and atom-readout pulses for atoms in the coupled (green) and uncoupled (orange) ground state manifolds, as well as for an empty cavity (gray). Transparency is proportional to occurrence, normalized to the highest of each case. Dotted lines are optimal threshold to minimize infidelity. (E) Measured infidelity for atom (pink) and state (purple) readout {\it vs.} probing time for two different probe powers. Black dotted curves are the theoretical expected infidelity from the known count-rates. Gray dotted lines are the measured best probing time in each case to find minimum fastest infidelity. (F) Lowest measured infidelity for atom (pink) and state (purple) readout {\it vs} probe count-rate, which is proportional to the probe power.}
\label{fig:readoutchar} 
\end{figure*}

Using time-tagged photon counts, we are able to extract an infidelity curve against integration time $\tau$ (Fig.~\refsub{fig:readoutchar}{E}). As expected from theory, the larger time we integrate the more separated both cases get as $\mu_L = C_{Atom}\tau$ and $\mu_H = C_M\tau$, and the lower the infidelity gets. However, we observe that for longer integration times, the infidelity stops decreasing. One observation is that atom-readout is always better than state-readout, which can be explained by unwanted off-resonant from the \SI{850}{\nano\meter} optical trap which state-readout does not correct for as there is no optical pumping. Even for scattering timescales on the order of \SI{100}{\milli\second}, at an integration time of \SI{10}{\micro\second} the infidelity is already limited to the order of $\sim10^{-4}$. The rest of the measured infidelity, we attribute to imperfect state preparation during the characterization experiment, which can be further optimized.

We can also look at the best measured infidelity as a measure of probe power, as shown in Fig.~\refsub{fig:readoutchar}{F}. At low powers, the count-rates are lower, so we need to integrate further. Therefore, the unwanted scattering becomes a more prominent limitation. At high powers, the weak-probe regime is broken and the effective ratio $C_M / C_{Atom}$ decreases, which sets the possible difference between High and Low counts and is proportional to the attainable readout fidelity. Therefore, we find an optimum at $C_P\sim\SI{1.75}{\mega\cps}$. We can, however, work at the stronger powers before the infidelity grows too large and have high-fidelity non-destructive state readout in the \SI{}{\micro\second} scale, limited by the effective dead-time of our detectors and our cavity linewidth $\kappa$.

\subsection{Error detection through Readout pulse sequences}
\subsubsection{Single atom Readout}
To read out the qubit states along with error states, we must conduct sequential readouts interspersed with coherent microwave rotations between our qubit states. We rely on our readout being non-destructive to the $\ket{0}$ state. Another important mechanism is that reading out an atom in the $\ket{1}$ state will quickly pumps it into the $\ket{2,2}$ stretched state. Our single-atom readout sequence entails two successive readouts with a microwave $\pi$ pulse applied in-between. Then, taking the above mechanisms into account, we find that the information from the two readout pulses will discriminate between the $\ket{0}$, $\ket{1}$ and error $\ket{err}$ states (Fig.~\refsub{fig:readoutseq}{A}). We denote a readout result as high (H) or low (L) depending on if the measured counts are above or below a selected threshold. This measurement determines whether an atom is in the coupled or uncoupled ground-state manifold. 

\subsubsection{Two atom Readout}
To extend this to two atoms, we perform three sequential readouts with the addition of a local pumping beam that pumps atom B into the uncoupled $F=1$ ground-state manifold, effectively “hiding” it from the readout. We note that when two atoms are coupled to the cavity, if either of the atoms is in the $\ket{1}$ state, the transmission of the cavity will be blocked. To distinguish which of the two atoms was in the coupled state, we must pump one of the two atoms out of the coupled state. To measure information about the two-atom state we first perform readout on both atoms, then hide atom B using a local depumping beam and then perform the single-atom readout sequence only on atom A.  This allows us to extract full information about the state of atom A, while also extracting some information about the two-atom state.

\begin{figure*}
\includegraphics[width=\textwidth]{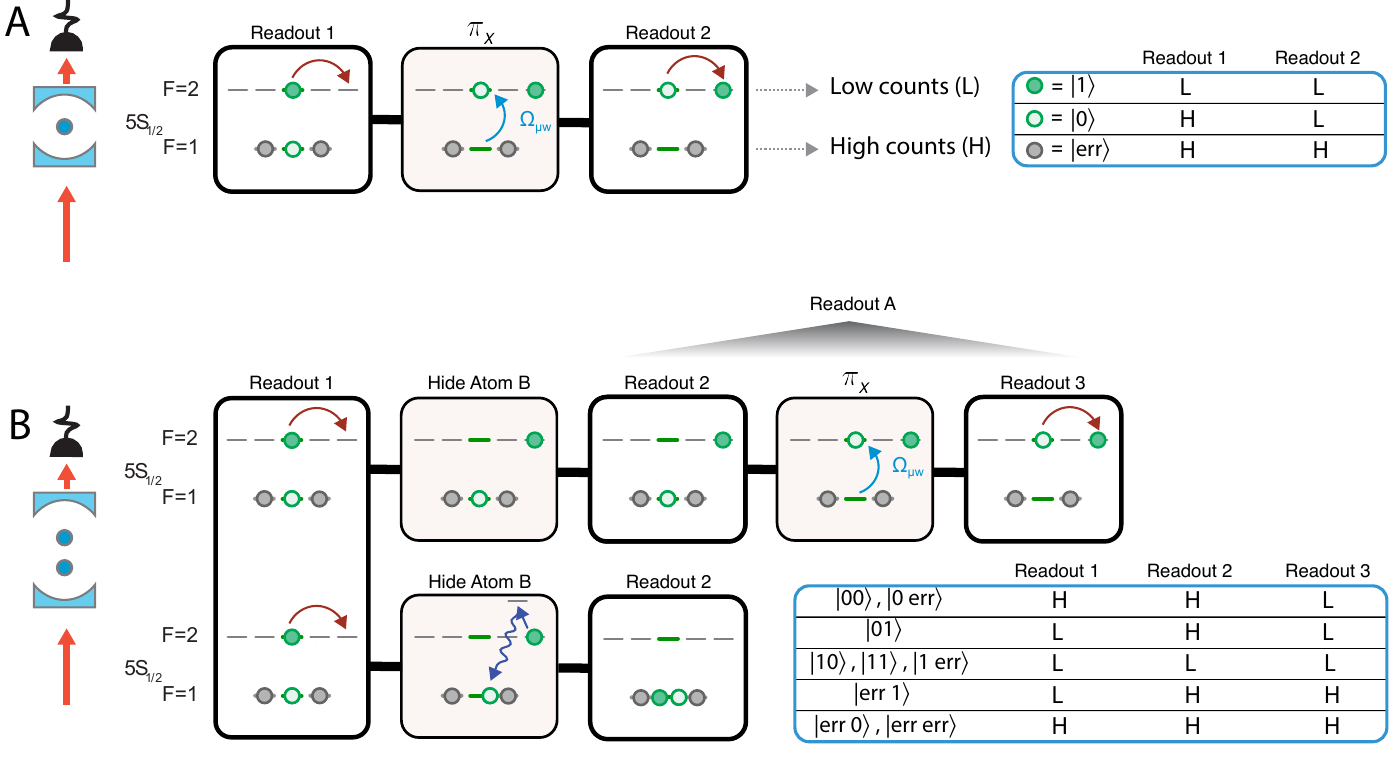}
\caption{Readout sequences for one (A) and two (B) atoms. The tables show the corresponding readout labels for each group of distinguishable states.} 
\label{fig:readoutseq} 
\end{figure*}

We find that this readout sequence can uniquely identify the populations in the $\ket{00}$ and $\ket{01}$ two-qubit states, while also being able to distinguish states where atom A has an error $\ket{err}_A$ (Fig.~\refsub{fig:readoutseq}{B}). We note that states where atom B is in the error state $\ket{err}_B$ are not distinguishable from $\ket{0}_B$. However, as explained in the entanglement generation section, these kinds of errors are suppressed.

\section{Cavity Spectra}
In the main text, we measure the spectra of the atom-cavity system under various driving conditions and number of atoms coupled to the cavity. The way that we do these measurements is in a “single-shot” as shown in Fig.~\refsub{fig:readoutseq}{A}. 
To understand the underlying physics of these spectra, we can consider the simplified Jaynes- and Tavis-Cummings Hamiltonians for one and two two-level atoms coupled to a single cavity mode. Using the rotating-wave approximation with respect to a drive, we can express the Hamiltonian in the rotating frame as:
\begin{align}
    \hat{H} / \hbar &= \Delta_c \hat{a}^\dagger\hat{a} \nonumber \\&+\sum_{\alpha=A,B}\left(\Delta_a\ket{e}\bra{e}_\alpha+g_\alpha\left(\ket{e}\bra{g}_\alpha\hat{a}+\ket{g}\bra{e}_\alpha\hat{a}^\dagger\right)\right)\label{eqn:Hamiltonian1}
\end{align}
Where $g_i$ is the single photon Rabi frequency of the atom $i$ coupled to the cavity. Furthermore, $\Delta_c$ and $\Delta_a$ are the cavity and atom detuning from the drive, respectively. The driving terms can be expressed as:
\begin{align}
    \hat{H}_{probe}/\hbar &= \Omega_{probe} \left(\hat{a}^\dagger + \hat{a}\right) \\
    \hat{H}_{side}/\hbar &=\sum_{\alpha=A,B} \frac{\Omega_{side,\alpha}}{2}\left(\ket{e}\bra{g}_\alpha+\ket{g}\bra{e}_\alpha\right) \nonumber\\
    &= \sum_{\alpha=A,B} \Omega_{side,\alpha}\hat{\sigma}_x^\alpha
\end{align}
Where $\hat{H}_{probe}$ is a drive of the cavity by probing it as we do for readout, and $\eta$ is the strength of said drive. Additionally, $\hat{H}_{side}$ considers a drive from the side of the cavity directly on the atoms, as we do for entanglement generation, where $\Omega_i$ is the strength of the side drive. 

\subsection{Weak-Probe Regime}
For the purposes of the spectra shown in Figs.~1-2 of the main text, assuming this simplified two-level system we can find an analytical form for the spectra assuming a steady-state solution, which we use to extract our cavity parameters. In the following, we expand on the derivation show in the supplement of~\cite{tiecke2014switch} accounting for multiple atoms and a side drive. From this treatment we arrive at equations for intracavity photon number for side and cavity probe given by:
\begin{align}
 \avg{a^\dagger a}_{probe} &\propto \abs{\frac{\Omega_{probe}\sqrt{\kappa}}{\left(1 + \tilde{C}_A + \tilde{C}_B\right)\tilde{\kappa}}}^2 \label{eqn:probespectra}\\ 
    \&\qquad \avg{a^\dagger a}_{side} &\propto \abs{\sqrt{\frac{\gamma}{\tilde{\kappa}\tilde{\gamma}}}\frac{\left(\Omega_{side,A}\sqrt{\tilde{C}_A} + \Omega_{side,B}\sqrt{\tilde{C}_B}\right)}{\left(1 + \tilde{C}_A + \tilde{C}_B\right)}}^2 \label{eqn:sidespectra}
\end{align}

In this steady-state solution, it stands to reason that the number of leaked photons over some integration time will be directly proportional to the number of photons inside the cavity. As a result, we use these analytical solutions to fit our cavity parameters to different spectra.

\subsection{Two-level Numerical Simulation}
An important result in our manuscript is the observation and use of photonic dark-states $\ket{D_0}$. These states can only be driven through a side drive, $\Omega_{side}$, directly pumping excitation on the atoms. This is because $\ket{D_0}$ states do not have any photonic component. However, to theoretically explain the measured spectra of our experiment, we cannot simply rely on the weak-probe regime analytical form of Eqn.~\ref{eqn:sidespectra}. This analytical form is sufficient to show the bright states $\ket{B_\pm}$ and $\sqrt{2}$ enhancement, but since the main approximation is that no atom ever gets excited, it cannot show any feature due to the photonic dark state.

To correctly account for full two atom spectra, we instead to solve for the steady-state using the master equation of the two-level atom Hamiltonian, shown in Eqn.~\ref{eqn:Hamiltonian1} using the python package QuTiP~\cite{qutip}. However, it is important to note that in the situation that the side drive $\Omega_{side}$ is on both atoms, the relative phase between the drive and coupling of each atom matters for whether we can excite either the bright or dark states. 
In practice, we always see the peaks corresponding to bright and dark states at the same time, as shown in Fig.~2 of the main text. We attribute this observation to thermal sampling of the atomic wave functions, leading to an effective incoherent drive of the system. For the purposes of fitting spectra, we build a fitting function that solves the master equation for Eqn.~\ref{eqn:Hamiltonian1} and averages both the case of in-phase and out-of-phase side drive. The fit is shown in Fig.~2D of the main text. Another interesting observation from these fits, is that they account for an effective “broadening” of the dark-state line due to strong $\Omega_{side}$ driving. 

\subsection{“Loss” Spectra}
One interesting observation of our entanglement generation protocol is the ability to conditionally couple atom A to a photonic dark state based on the state of atom B (Fig.~3A of main text). Particularly, in the carving scheme we quickly pump state $\ket{10}$ into the detectable error states $\ket{err}_A$ while other states like $\ket{11}$ are pumped much slower. This asymmetry in their decay, which is shown in Fig.~3B of the main text, is what allows us to carve out parts of the state and generate high fidelity bell-states. However, it does depend on the side drive on atom A being exactly in resonance with the photonic dark state in the $2\leftrightarrow1'$ transition. We calibrate this detuning by measuring what we call “loss” spectra.

We can understand the “loss” spectra as simply a measurement of the probability of being in the error state $\ket{err}_A$ after a pulse of some short length. For example, we can choose a pulse length from Fig.~3B where state $\ket{10}$ has not been completely pumped into the error state and vary the probe detuning.

\begin{figure*}
\centering
\includegraphics[width = \textwidth]{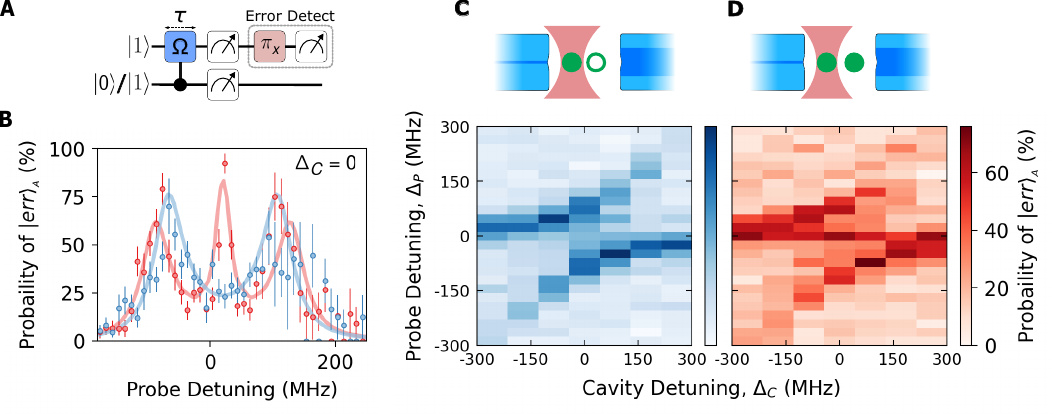}
\caption{(A) Quantum circuit for preparing two-qubit states and measuring the $\ket{err}_A$ population. (B) With the cavity tuned to the $2\leftrightarrow1'$ transition, we measure the probability of transferring population to the $\ket{err}_A$ state as a function of probe detuning for the $\ket{11}$ state (blue) and the $\ket{10}$ state (red). (C-D) Measurements of the probability of transferring population to the $\ket{err}_A$ as a function of both probe detuning and cavity detuning for the $\ket{11}$ state (C) and the $\ket{10}$ state (D)}
\label{fig:fullspectra} 
\end{figure*}
\section{Experimental Characterization of the Entanglement Fidelity} 
\subsection{Measuring Entanglement Fidelity}
In our experiments, we use two different entanglement generation schemes to prepare two-qubit bell states of the form $\ket{\Phi^\pm}=\frac{1}{\sqrt{2}}\left(\ket{00} \pm \ket{11}\right)$. If we assume that after an experiment, our two-atom state is described by a density matrix $\rho$, then the entanglement fidelity is simply:
\begin{align}
    \mathcal{F}_{\ket{\Phi^\pm}} = \bra{\Phi^\pm}\rho\ket{\Phi^\pm} = \frac{1}{2}\left(\rho_{00,00}\pm\rho_{00,11}\pm\rho_{11,00}+\rho_{11,11}\right)
\end{align}

As explained in the main text, we measure the diagonal elements of the quantum state in all three possible bases XX, YY, and ZZ by controlling the relative phase of global microwave pulses and rotating state $\rho$. We measure the population in $\ket{11}$ and $\ket{10}$ by performing a $\pi$-pulse for a total of 6 measurements: 2 measurements per each basis. This gives us a total of 12 data points corresponding to the 4 diagonal populations in each basis (i.e., $P_z(\ket{00}) = P_{00}^z=\rho_{00,00}$). The ZZ basis is the one in which we define our state. The XX and YY basis are accesses using $\pi/2$ rotations into the equator of the generalized Bloch sphere. With our measurements, we have direct access to both the parity of the state and its trace as measured in different basis:
\begin{align}
    \Pi_i &= P_{00}^i - P_{01}^i - P_{10}^i + P_{11}^i \label{eqn:parity}\\
    \text{Tr}_i & = P_{00}^i + P_{01}^i + P_{10}^i + P_{11}^i \label{eqn:trace}
\end{align}
Where $i=x,y,z$ is the measurement basis,  $\Pi_i$ is the state parity in basis $i$, and $\text{Tr}_i$ is the measure trace over the qubit manifold in basis $i$. Importantly, since the qubit rotations do not take us out of the qubit manifold, it is easy to show that:
\begin{align}
    \text{Tr}_x=\text{Tr}_y=\text{Tr}_z=\rho_{00,00}+\rho_{01,01}+\rho_{10,10}+\rho_{11,11}
\end{align}
In the case where there are no errors, $\text{Tr}_i = 1$, as expected for a valid density matrix $\rho$. However, this is not the case of states which have component in error states outside the qubit manifold as in the case of our deterministic entanglement measurement.

Furthermore, the parities in the different bases can be expressed as:
\begin{align}
    \Pi_x &= -\rho_{00,11}+\rho_{01,10}+\rho_{10,01}-\rho_{11,00} \\
    \Pi_y &= +\rho_{00,11}+\rho_{01,10}+\rho_{10,01}+\rho_{11,00} \\
    \Pi_z &= +\rho_{00,00}-\rho_{01,01}-\rho_{10,10}+\rho_{11,11}
\end{align}
This relations let us rewrite the entanglement fidelity in terms of the trace and state parity as:
\begin{align}
    \mathcal{F}_{\ket{\Phi^\pm}} = \frac{1}{4}\left(\avg{\text{Tr}} \pm \Pi_x \mp \Pi_y + \Pi_z\right)
    \label{eqn:fidelity}
\end{align}
Where $\avg{\text{Tr}} = \frac{1}{3}\left(\text{Tr}_x + \text{Tr}_y + \text{Tr}_z\right)$, is the average trace over all basis to remove any ambiguity related to the finite sampling size or imperfect microwave rotations. The sign for the X and Y parities reflects what is expected for each bell-state and is explicitly shown in the choice of direction for the measured correlations in Fig.~3C and Fig.~4D of the main text. To extract the entanglement fidelity, we substitute the values in Eqn.~\ref{eqn:fidelity}, using the expressions for the parity (Eqn.~\ref{eqn:parity}) and trace (Eqn.~\ref{eqn:trace}) in terms of the 12 measured values and propagate uncertainty assuming uncorrelated errors. This equation has the benefit that it is agnostic to whether or not we are using error detection in our measurement, and is how we are able to measure verifiable deterministic entanglement.

\section{Theoretical Models}
\subsection{First excitation manifold}
Our entanglement scheme relies on the fact that the strong cooperativity in our system, allows us to spectroscopically resolve a photonic two-atom dark-state from all other possible states. This capability enables conditional operations using a ground-state qubit encoding between cavity coupled $\ket{1}$ and uncoupled $\ket{0}$ states. 

Our scheme can be simply understood considering the simplified two-level atom system described by the Tavis-Cummings Hamiltonian shown in Eqn.~\ref{eqn:Hamiltonian1}. When both the cavity, atom, and drive are resonant such that $\Delta=\Delta_a=\Delta_c=0$, it is easy to diagonalize the Hamiltonian and find the eigenvalues of the single-excitation manifold. In the case of only atom A coupled to the cavity where $g_B = 0$ (i.e., Atom B is in the uncoupled state $\ket{0}$, the eigenstates take the form:
\begin{align}
    \ket{\pm} = \frac{1}{\sqrt{2}}\left(\ket{e,0}\pm\ket{g,1}\right)
\end{align}
With their eigenvalues being $E_\pm=\pm g_A$. This strong vacuum-Rabi splitting effectively suppresses the drive excitation rate proportionally to the single-atom cooperativity $C$.

However, when both atoms are equally strongly coupled to the cavity $g_A=g_B=g_0$, there are now three eigenstates:
\begin{align}
    \ket{B_\pm} &= \frac{1}{2} \left( \ket{eg,0} + \ket{ge,0}\right) \pm \frac{1}{\sqrt{2}}\ket{gg,1} \\
    \ket{D_0} &= \frac{1}{2} \left( \ket{eg,0} - \ket{ge,0}\right) 
\end{align}
The two “bright” states $\ket{B_\pm}$ are analogous to the single atom eigenstates with a stronger repulsion due to it being two atoms  $E_\pm=\sqrt{2}g_0$. We denote these states as “bright” states due to their cavity-like components that allow them to decay through the cavity. In contrast, we also find a third zero energy eigenstate $\ket{D_0}$. We label this a photonic “dark” state due to the lack of a cavity excitation. One can understand this state as a sub-radiant state in which the emission from the two atoms destructively interferes with each other’s emission into the cavity, resulting in a state that cannot decay through the cavity. 

\subsection{Carving protocol}
Now we can consider a more complicated level structure in which one of the atom, atom A, has a level structure composed of two qubit states, $\ket{0}$ and $\ket{1}$, an excited state, $\ket{e}$, and a state we will denote as the error state, $\ket{err}$. The $\ket{err}\rightarrow\ket{e}$ transition is coupled to the cavity, while the qubit states are not. Atom B will have similar structure with qubit states, $\ket{0}$ and $\ket{1}$, and the excited state, $\ket{e}$, in this atom the $\ket{0}\rightarrow\ket{e}$ state will be coupled to the cavity. The gate can be performed with a global beam; however, this requires a well-defined phase of the drive between the two qubits. To circumvent this issue, we perform the gate by driving only atom A on the $\ket{1}\rightarrow\ket{e}$ transition.

Now with qubit states we can consider the evolution of each state. For the states where atom A is in $\ket{0}$ there is no evolution, the interesting cases are when atom A is in state $\ket{0}$. For the initial state, $\ket{10}$, the transition to the excited state will be allowed through this two-atom dark state due to the second atom being in a state coupled to the cavity, $\ket{0}$. For the initial state $\ket{11}$, the evolution will be the same as if only atom A were coupled to the cavity, resulting in a blockade of the excitation. If we treat this case more carefully, we will find a single atom dark state that is formed between the states given by: 
\begin{align}
    \frac{g}{\sqrt{\Omega^2+g^2}}\ket{1}\ket{0}_{ph}-\frac{\Omega}{\sqrt{\Omega^2+g^2}} \ket{err}\ket{1}_{ph}
\end{align}
This will result in decay through the cavity into the $\ket{err}\ket{0}_{ph}$ state with a rate given by:
\begin{align}
    \gamma_{\ket{11}} &= \frac{\Omega^2}{\Omega^2 + g^2}\kappa\approx \frac{\Omega^2}{C\gamma}
\end{align}
Now we investigate the $\ket{10}$  case. We can approximate the dynamics of this state by assuming $g>\Omega$ thus the drive is a small perturbation coupling the $\ket{10}$  state to the dark state, $\ket{E}=\frac{1}{\sqrt{2}}\left(\ket{e,0}-\ket{err,e}\right)$. The drive term takes the form of $H_{dr} = \Omega_A\ket{e}\bra{g}_A+\Omega_A^* \ket{g}\bra{e}_A$. Under these approximations we can assume that the rotation between an effective two-level system with the ground state, $\ket{G}=\ket{10}$ and excited state, |E⟩ with a transition dipole matrix element given by:
\begin{align}
    \hat{H}_{eff} &= \frac{\Omega}{\sqrt{2}}\ket{E}\bra{G} +  \frac{\Omega^*}{\sqrt{2}}\ket{G}\bra{E}
\end{align}
We find that the dark state will decay at the same rate as an atom in free space, $\gamma$. As a result, the drive will depopulate the $\ket{10}$  state at a rate given by:
\begin{align}
    \gamma_{\ket{10}}=\frac{\Omega_{eff}^2}{\gamma} = \frac{\Omega^2}{2\gamma}
\end{align}

The excitation of the dark state can decay through atom A or atom B with a $50\%$ probability due to the equal superposition of excitation. If the decay were to be through atom B, this would trap atom A in the error state, resulting in a fully detectable error. If atom A were to decay, the population will not always go to the error states, this decay will be determined by the branching ratios of the excited state. Through careful selection of atomic states, this ratio can be large and allow for almost fully detectable errors. This branching ratio sets a limit on the fidelity that one can achieve with this approach. 

Due to the large ratio of decay between the $\ket{10}$ and $\ket{11}$ states, we can pump one state into the error state much faster than the other. Upon post selection of not being in the error state, this will then “carve out” one of the two states. We model this as applying a matrix:
\begin{align}
  \label{eqn1}
    \hat{U}_g &= \begin{pmatrix}
        1 & 0 & 0 & 0 \\ 
        0 & 1 & 0 & 0 \\ 
        0 & 0 & e^{-\gamma_{\ket{10}}t/2} & 0 \\
        0 & 0 & 0 & e^{-\gamma_{\ket{11}}t/2} 
    \end{pmatrix}
\end{align}
Applying a $\pi$ pulse in between two of these pulses will result in a matrix:
\begin{align}
    \hat{U}_{carve} &= \hat{U}_g \hat{R}_x(\pi) \hat{U}_g \nonumber\\ &=\begin{pmatrix}
        e^{-\gamma_{\ket{11}}t/2} & 0 & 0 & 0 \\ 
        0 & e^{-\gamma_{\ket{10}}t/2} & 0 & 0 \\ 
        0 & 0 & e^{-\gamma_{\ket{10}}t/2} & 0 \\
        0 & 0 & 0 & e^{-\gamma_{\ket{11}}t/2} 
    \end{pmatrix}
\end{align}
By applying this matrix and renormalizing the population, we find that fidelity will exponentially approach unity while the success rate linearly decreases for large cooperativities.
\begin{equation}
    P_{succ} = \bra{\Psi_0}\hat{U}_{carve}^\dagger\hat{U}_{carve}\ket{\Psi_0} = \frac{e^{-\frac{\Omega^2}{C\gamma}t} + e^{-\frac{\Omega^2}{2\gamma}t}}{2} 
\end{equation}
\begin{align}
    \mathcal{F} &= \abs{\frac{\bra{\Phi^+}\hat{U}_{carve}\ket{\Psi_0}}{\sqrt{\bra{\Psi_0}\hat{U}_{carve}^\dagger\hat{U}_{carve}\ket{\Psi_0}}}}^2 \nonumber\\
    &= \frac{e^{-\frac{\Omega^2}{C\gamma}t}}{e^{-\frac{\Omega^2}{C\gamma}t} + e^{-\frac{\Omega^2}{2\gamma}t}} =\frac{e^{-\frac{\Omega^2}{C\gamma}t}}{2 P_{succ}} 
\end{align}

\subsection{Simplified Two-Level Model for CZ Gate}
Using the same setup as the carving procedure, we can also perform a controlled phase gate. The difference between the carving case and this case is that we drive with a stronger field, such that one performs a full $2\pi$ rotation between the ground state and the dark state. This understanding of the gate assumes that we are in a regime where the dark state is defined well with respect to the strength of the drive and that the drive is fast enough to perform a coherent $2\pi$ rotation. Overall, this results in the condition $g>\Omega>\gamma$. The scattering off the dark state will determine the success rate of the gate. 
\begin{align}
\label{eq:resonant}
    \epsilon_{\ket{10}} = 1 - e^{-\frac{\pi\gamma}{2\Omega}}\approx \frac{\pi\gamma}{2\Omega}
\end{align}
The fidelity will then be limited to how much of this scattering results in a detectable error. decay out of the second atom will be completely detectable because the excitation of the second atom is correlated with the first atom being in the error state. For the other part of the wave function, we consider decay to states with undetectable errors, for example, decaying back into the qubit manifold. We can denote this decay as $\gamma_{bad}$. This results in a total uncorrectable error, given by:
\begin{align}
\epsilon_{uncorr}\approx\frac{\gamma_{bad}}{\gamma}\frac{\pi\gamma}{2\Omega}
\end{align}
For now, we will consider $\gamma_{bad}=0$ and see how the fidelity will scale. We can consider the case with the initial state $\ket{11}$. For this state we can treat the system as if it is a single atom coupled to the cavity as shown in the carving protocol. Given the gate time, $\frac{\pi}{\Omega}$, this will result in a decay of the $\ket{11}$ state into the $\ket{err}_A$ state of:
\begin{align}
\label{eq:suppressed}
    \epsilon_{\ket{11}} = 1- e^{-\frac{\pi\gamma_{eff}}{\Omega}}\approx\frac{\pi\Omega\kappa}{\Omega^2+g_{1-1'}^2} \approx \frac{\pi\Omega}{C\gamma}
\end{align}

In principle, all these errors should be detectable. Now, we must consider what correction does to our final state. The unitary that the gate will apply is given below. Here we do not include the error state manifold making this matrix not conserve population:
\begin{align}
    \hat{U}_g &= \begin{pmatrix}
        1 & 0 & 0 & 0 \\ 
        0 & 1 & 0 & 0 \\ 
        0 & 0 & -e^{-\frac{\pi\gamma}{2\Omega}} & 0 \\
        0 & 0 & 0 & e^{-\frac{\pi\Omega\kappa}{\Omega^2+g_{1-1'}^2}} 
    \end{pmatrix}
\end{align}
If we consider the ideal CZ gate to be:
\begin{align}
    \hat{U}_{CZ} &= \begin{pmatrix}
        1 & 0 & 0 & 0 \\ 
        0 & 1 & 0 & 0 \\ 
        0 & 0 & -1 & 0 \\
        0 & 0 & 0 & 1 
    \end{pmatrix}
\end{align}
Then we can define a gate fidelity as: 
\begin{align}
    \mathcal{F} &= |{\bra{\Psi_0}\hat{U}_{CZ}^\dagger\hat{U}_{g}\ket{\Psi_0}}|^2 \nonumber \\
    &= \frac{1}{16}\left( 2 +  e^{-\frac{\pi\gamma}{2\Omega}} + e^{-\frac{\pi\Omega\kappa}{\Omega^2+g_{1-1'}^2}} \right)^2
\end{align}

Given that we are in the strong coupling regime we can use the approximations from \ref{eq:resonant} and \ref{eq:suppressed} to simplify this equation. We can then maximize fidelity by finding an optimal Rabi frequency. We find an optimal Rabi frequency and maximum fidelity, given by:
\begin{align}
    \Omega_{opt} &= \sqrt{\frac{g_{1-1'}^2}{\frac{2\kappa}{\gamma}-1}}\approx\sqrt{\frac{C}{2}}\gamma \\
    \mathcal{F}_{uncorr} &\approx \left(1 - \frac{\pi}{2\sqrt{2C}}\right)^2
\end{align}
We find that the optimal fidelity is found when the loss of the $\ket{10}$ and $\ket{11}$ are symmetric. After post-selection, the state will be projected out of the error state using the projection matrix $\hat{M}=1-\ket{err}\bra{err}$. Since the gate operator $\hat{U}_g$ is not population conserving, we can estimate the fully corrected gate fidelity as:
\begin{align}
    \mathcal{P}_{success} &= \bra{\Psi_0}\hat{U}_g^\dagger\hat{U}_g\ket{\Psi_0} = 1-\frac{\pi}{\sqrt{2C}}\\
    \mathcal{F}_{corr} &= \frac{\mathcal{F}_{uncorr}}{ \mathcal{P}_{success}} = \frac{\left(1 - \frac{\pi}{2\sqrt{2C}}\right)^2}{1-\frac{\pi}{\sqrt{2C}}}
\end{align}

We find that the fidelity is limited by the correction creating an imbalance of population between states where atom A is in $\ket{0}_A$ and states where atom A is in $\ket{1}_A$. By introducing a rotation, $\alpha$, between the $\ket{0}_A$ and $\ket{err}_A$ we can improve the fidelity of the gate. By including this operation in the gate and using the results for the optimal drive strength above, we find that there exists a value for $\alpha$ such that the corrected gate fidelity is unity at the cost of success probability. This $\alpha$ can be intuitively understood as matching the loss of all qubit states:
\begin{align}
    \alpha &= e^{-\frac{\pi}{\sqrt{2C}}}\\
    \mathcal{F}_{uncorr} = \mathcal{P}_{success}  &= e^{-\frac{2\pi}{\sqrt{2C}}}\approx 1-\frac{2\pi}{\sqrt{2C}}\\
    \mathcal{F}_{corr} &= \frac{\mathcal{F}_{uncorr}}{ \mathcal{P}_{success}} = 1
\end{align}

\section{Theoretical Modelling for Rb-87}

\begin{figure}[h]
\centering
\includegraphics[width = 0.4\textwidth]{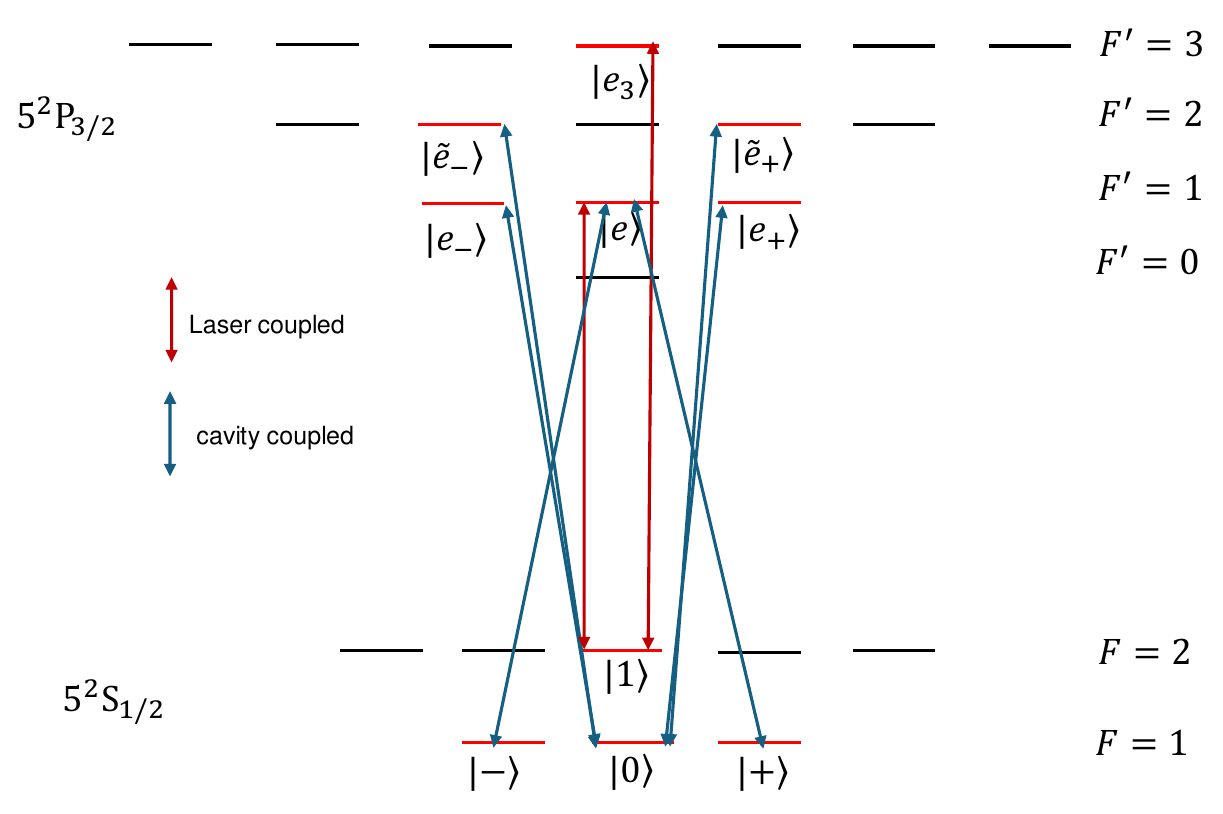}
\caption{Hyperfine structure of the $\text{D}_2$ line in $^{87}$Rb. The laser driving and cavity field couples only a small subset of the hyperfine states marked in red under the assumptions mentioned in the main text.} 
\label{fig:hyperfine} 
\end{figure}

The full hyperfine structure of the $^{87}$Rb $\text{D}_2$ line is shown in Fig.~\ref{fig:hyperfine}. We assume negligible Zeeman splitting, such that all states in the same $F$ ($F'$) manifold are degenerate in energy and that we are operating on two-photon resonance such that $\omega_c-\omega_L+\omega_{0}-\omega_1=0$, where $\omega_c$ is the cavity resonance frequency, $\omega_L$ is the laser frequency, and $\omega_0$ ($\omega_1$) is the frequency of qubit state $\ket{0}$ ($\ket{1}$). Under these assumptions, only a small subset of states will be coupled by the laser driving and cavity field when both atoms start out in the qubit manifold and only atom 1 is driven. This allows us to write down the Hamiltonian of the two-atom system as
\begin{align} \label{eq:hamil1}
\hat{H}=&\Delta(\ket{5}\bra{5}+\ket{6}\bra{6}+\ket{11}\bra{11})+\Delta_3(\ket{7}\bra{7}+\ket{8}\bra{8}) \nonumber \\
&+\Delta_2\ket{12}\bra{12}+(\Omega(\ket{5}\bra{3}-\sqrt{9}\ket{7}\bra{3}+\ket{6}\bra{4}\nonumber\\
&-\sqrt{9}\ket{8}\bra{4})+h.c.)+(g(\ket{11}\bra{9}+\sqrt{2}\ket{5}\bra{9}\nonumber\\
&+\sqrt{2}\ket{6}\bra{10}+\sqrt{\frac{1}{8}}\ket{12}\bra{9}) +h.c.)
\end{align}
as well as Lindblad operators describing spontaneous emission ($\gamma$) and cavity decay ($\kappa$):
\small
\begin{equation} \label{eq:lindblad1}
\begin{aligned}
\hat{L}_{c1}&=\sqrt{\frac{\kappa}{2}}(\ket{13}\bra{9}+\ket{14}\bra{10}),\qquad \\
\hat{L}_{d}&=\sqrt{\gamma_d}(\ket{17}\bra{5}+\ket{18}\bra{6}), \qquad\\
\hat{L}_{3d}&=\sqrt{\gamma_{3d}}(\ket{21}\bra{7}+\ket{22}\bra{8}),\qquad \\
\hat{L}_{e1}&=\sqrt{\frac{\gamma}{2}}\ket{25}\bra{11},\qquad \\
\hat{L}_{\tilde{e}1}&=\sqrt{\frac{\gamma}{2}}\ket{27}\bra{12}, \qquad
\end{aligned}
\begin{aligned}
\hat{L}_{c2}&=-\sqrt{\frac{\kappa}{2}}(\ket{15}\bra{9}+\ket{16}\bra{10}), \\
\hat{L}_{nd}&=\sqrt{\gamma_{nd}}(\ket{19}\bra{5}+\ket{20}\bra{6}), \\
\hat{L}_{3nd}&=\sqrt{\gamma_{3nd}}(\ket{23}\bra{7}+\ket{24}\bra{8}), \\
\hat{L}_{e2}&=\sqrt{\frac{\gamma}{2}}\ket{26}\bra{11}, \\
 \hat{L}_{\tilde{e}2}&=\sqrt{\frac{\gamma}{2}}\ket{28}\bra{12}.
\end{aligned}
\normalsize
\end{equation}
Here, we have defined the following basis 
\begin{eqnarray}
&&\big\{\ket{0,0}\ket{0,0}_{c},\ket{0,1}\ket{0,0}_{c},\ket{1,0}\ket{0,0}_{c},\ket{1,1}\ket{0,0}_{c},\nonumber \\
&&\ket{e,0}\ket{0,0}_{c},\ket{e,1}\ket{0,0}_{c},\ket{e_3,0}\ket{0,0}_{c},\ket{e_3,1}\ket{0,0}_{c}, \nonumber \\
&&\ket{S_0},\ket{S_1},\ket{S_{e}},\ket{\tilde{S}_{e}},\ket{-,0}\ket{0,0}_c,\ket{-,1}\ket{0,0}_c,\nonumber \\
&&\ket{+,0}\ket{0,0}_c,\ket{+,1}\ket{0,0}_c,\ket{O_{d},0}\ket{0,0}_c, \ket{O_{d},1}\ket{0,0}_c,  \nonumber \\
&&\ket{O_{nd},0}\ket{0,0}_c,\ket{O_{nd},1}\ket{0,0}_c,\ket{O_{3d},0}\ket{0,0}_c,\nonumber \\
&&\ket{O_{3d},1}\ket{0,0}_c,\ket{O_{3nd},0}\ket{0,0}_c,\ket{O_{3nd},1}\ket{0,0}_c,\nonumber \\
&&\ket{-,O_{1}}\ket{0,0}_c,\ket{+,O_{2}}\ket{0,0}_c,\ket{-,\tilde{O}_{1}}\ket{0,0}_c,\ket{+,\tilde{O}_{2}}\ket{0,0}_c\big\}  \nonumber \\
&&=\{\ket{1},\ldots,\ket{28}\}.
\end{eqnarray}
Here the notation $\ket{x,y}\ket{1,0}_{c}$ denotes atom 1 (2) in state $\ket{x}$ ($\ket{y}$) with one (zero) $\sigma_+$ ($\sigma_-$) polarized cavity photon. Additionally, we have defined the states 
\begin{eqnarray}
\ket{S_x}&=\frac{1}{\sqrt{2}}\left(\ket{-,x}\ket{1,0}_c -\ket{+,x}\ket{0,1}_c\right) \\
\ket{S_{e}}&=\frac{1}{\sqrt{2}}\left(\ket{-,e_{+}}\ket{0,0}_c +\ket{+,e_{-}}\ket{0,0}_c\right) \\
\ket{\tilde{S}_{e}}&=\frac{1}{\sqrt{2}}\left(\ket{-,\tilde{e}_{+}}\ket{0,0}_c -\ket{+,\tilde{e}_{-}}\ket{0,0}_c\right),
\end{eqnarray}
and the decay states $\ket{O_{d}}$ ($\ket{O_{nd}}$) to capture spontaneous emission to detectable (non-detectable) states from state $\ket{e}$ and similarly $\ket{O_{3d}}$ ($\ket{O_{3nd}}$) for state $\ket{e_3}$. Note that $\gamma_{nd}+\gamma_{d}=\gamma_{3nd}+\gamma_{3d}=\gamma$, where $\gamma$ is the natural linewidth of the D2 transition. The ratio between the different decay rates into detectable and non-detectable states can be derived from the Clebsch-Gordan coefficients of the corresponding transitions. The decay states $\ket{O_1},\ket{O_2},\ket{\tilde{O}_1}, \ket{\tilde{O}_2}$ are introduced to capture decay from the excited states $\ket{S_e}$ and $\ket{\tilde{S}_e}$ in atom 2, which is always detectable since it leaves atom 1 in state $\ket{\pm}$. This simplified model neglects additional higher-order dynamics following spontaneous emission from the atoms, which should be a valid approximation in the regime of weak driving and strong cooperativity considered here. 
\subsection{Carving protocol}
For the carving protocol, the laser drive is weak ($\Omega\ll \gamma, g$), which allows us to adiabatically eliminate the entire excited subspace of the two-atom cavity system. Using the effective operator formalism from Ref.~\cite{Reiter2012}, we can analytically derive an effective Hamiltonian and effective Lindblad operators describing the evolution of the qubit ground states. We find that these are of the form:
\begin{eqnarray}
\hat{H}_{\text{eff}}&=&-\tilde{\Delta}_3\ket{3}\bra{3}-\tilde{\Delta}_{4}\ket{4}\bra{4}, \\
\hat{L}_{3d}&=&\sqrt{\Gamma_{3d}}\ket{d}\bra{3}, \\
\hat{L}_{3nd}&=&\sqrt{\Gamma_{3nd}}\ket{nd}\bra{3}, \\
\hat{L}_{4d}&=&\sqrt{\Gamma_{4d}}\ket{d}\bra{4}, \\
\hat{L}_{4nd}&=&\sqrt{\Gamma_{4nd}}\ket{nd}\bra{4},
\end{eqnarray}
where we have introduced the notation $\ket{d}$ ($\ket{nd}$) to denote a non-specified subspace of detectable (non-detectable) error states.  Driving on resonance ($\Delta=0$), we can simplify the full analytical expressions in the large cooperativity limit ($C\gg1$) to obtain
\begin{eqnarray}
\tilde{\Delta}_3&\approx& \left(\frac{9}{\Delta_3}+\frac{1}{36\Delta_2}\right)\Omega^2 \\
\tilde{\Delta}_4&\approx& \frac{9\Omega^2}{\Delta_3} \\
\Gamma_{3d}&\approx&\frac{8\gamma+4\gamma_d}{9}\frac{\Omega^2}{\gamma^2} \\ 
\Gamma_{3nd}&\approx&\frac{4\gamma_{nd}}{9}\frac{\Omega^2}{\gamma^2} \\
\Gamma_{4d}&\approx&\frac{\Omega^2}{2\gamma^2 C}\gamma+\frac{9\Omega^2}{\Delta_3^2}\gamma_{3d} \\
\Gamma_{4nd}&\approx&\frac{\Omega^2}{16\gamma^2 C^2}\gamma_{nd}+\frac{9\Omega^2}{\Delta_3^2}\gamma_{3nd},
\end{eqnarray}
where we have also taken the limit of $\Delta_3,\Delta_2\gg\gamma$. 

For the carving protocol, the atoms are first prepared in the state $(\ket{0,0}+\ket{0,1}+\ket{1,0}+\ket{1,1})/2$ and a laser drive is applied to atom 1 for a time $t$. After this, the qubit states are flipped and the same pulse is applied. Assuming that the single qubit operations are perfect and that all detectable error states are perfectly detected, the atoms will be prepared in a state
\begin{align}
\rho=&\frac{1}{P}(\ket{\psi}\bra{\psi}+\frac{1}{2}(\frac{\Gamma_{4nd}}{\Gamma_4}(1-e^{-\Gamma_4t})\nonumber\\
&+\frac{\Gamma_{3nd}}{\Gamma_3}(1-e^{-\Gamma_3t}))\ket{nd}\bra{nd}),
\end{align}
where 
\begin{align}
\ket{\psi}=&\frac{1}{2}( e^{(i\tilde{\Delta}_4-\Gamma_4/2)t}\ket{0,0}+e^{(i\tilde{\Delta}_3-\Gamma_3/2)t}\ket{0,1}\nonumber\\
&+e^{(i\tilde{\Delta}_3-\Gamma_3/2)t}\ket{1,0}+e^{(i\tilde{\Delta}_4-\Gamma_4/2)t}\ket{1,1} )
\end{align}

and $\Gamma_3=\Gamma_{3d}+\Gamma_{3nd}$, $\Gamma_4=\Gamma_{4d}+\Gamma_{4nd}$. The success probability (i.e. probability not to detect the atoms in an error state) is
\begin{equation}
P=\frac{\Gamma_4\Gamma_3-\frac{1}{2}(\Gamma_{4d}\Gamma_3+\Gamma_4\Gamma_{3d})}{\Gamma_3\Gamma_4}+\frac{1}{2}\left(\frac{\Gamma_{4d}}{\Gamma_4}e^{-\Gamma_4 t}+\frac{\Gamma_{3d}}{\Gamma_3}e^{-\Gamma_3 t}\right).
\end{equation}
The fidelity with the target Bell state $\ket{\phi}=(\ket{0,0}+\ket{1,1})/\sqrt{2}$ is $F=\bra{\phi}\rho\ket{\phi}=\frac{1}{2P}e^{-\Gamma_4 t}$.

\subsection{CZ-gate}

For the CZ-gate, the laser drive is stronger, and we are in the regime $g\gg\Omega\gg\gamma$. We can therefore not directly adiabatically eliminate the entire excited manifold, as in the case of the carving protocol. Nonetheless, the dynamics can be treated separately depending on whether the initial atomic state is $\ket{1,0}$ ($\ket{3}$) or $\ket{1,1}$ ($\ket{4}$). Specifically, we can write the Hamiltonian in Eqn.~(\ref{eq:hamil1}) as $\hat{H}=\hat{H}_3+\hat{H}_4$, where

\begin{align}
\hat{H}_3=&\Delta(\ket{5}\bra{5}+\ket{11}\bra{11})+\Delta_3\ket{7}\bra{7}\nonumber\\
&+\Delta_2\ket{12}\bra{12}+\left(\Omega(\ket{5}\bra{3}-\sqrt{9}\ket{7}\bra{3})+h.c.\right)\nonumber\\
&+\left(g(\sqrt{2}\ket{5}+\ket{11}+\sqrt{1/8}\ket{12})\bra{9}+h.c.\right)
\end{align}

\begin{align}
\hat{H}_4=& \Delta \ket{6}\bra{6}+\Delta_3\ket{8}\bra{8}\nonumber\\
&+\left(\Omega(\ket{6}\bra{4}-\sqrt{9}\ket{8}\bra{4})+h.c.\right)\nonumber\\
&+\sqrt{2}\left(g\ket{6}\bra{10}+h.c.\right),
\end{align}
and consider the evolution under each Hamiltonian separately. The corresponding Lindblad operators for the two subsystems are
\begin{equation} \label{eq:lindbladg3}
\begin{aligned}
\hat{L}^{(3)}_{c1}&=\sqrt{\frac{\kappa}{2}}(\ket{13}\bra{9}),\qquad \\
\hat{L}^{(3)}_{d}&=\sqrt{\gamma_d}(\ket{17}\bra{5}), \qquad\\
\hat{L}^{(3)}_{3d}&=\sqrt{\gamma_{3d}}(\ket{21}\bra{7}),\qquad \\
\hat{L}^{(3)}_{e1}&=\sqrt{\frac{\gamma}{2}}\ket{25}\bra{11},\qquad \\
\hat{L}^{(3)}_{\tilde{e}1}&=\sqrt{\frac{\gamma}{2}}\ket{27}\bra{12}, \qquad
\end{aligned}
\begin{aligned}
\hat{L}^{(3)}_{c2}&=-\sqrt{\frac{\kappa}{2}}(\ket{15}\bra{9}), \\
\hat{L}^{(3)}_{nd}&=\sqrt{\gamma_{nd}}(\ket{19}\bra{5}), \\
\hat{L}^{(3)}_{3nd}&=\sqrt{\gamma_{3nd}}(\ket{23}\bra{7}), \\
\hat{L}^{(3)}_{e2}&=\sqrt{\frac{\gamma}{2}}\ket{26}\bra{11}, \\
 \hat{L}^{(3)}_{\tilde{e}2}&=\sqrt{\frac{\gamma}{2}}\ket{28}\bra{12},
\end{aligned}
\end{equation}
and 
\begin{equation} \label{eq:lindbladg4}
\begin{aligned}
\hat{L}^{(4)}_{c1}&=\sqrt{\frac{\kappa}{2}}(\ket{14}\bra{10}),\qquad \\
\hat{L}^{(4)}_{d}&=\sqrt{\gamma_d}(\ket{18}\bra{6}), \qquad\\
\hat{L}^{(4)}_{3d}&=\sqrt{\gamma_{3d}}(\ket{22}\bra{8}),\qquad
\end{aligned}
\begin{aligned}
\hat{L}^{(4)}_{c2}&=-\sqrt{\frac{\kappa}{2}}(\ket{16}\bra{10}), \\
\hat{L}^{(4)}_{nd}&=\sqrt{\gamma_{nd}}(\ket{20}\bra{6}), \\
\hat{L}^{(4)}_{3nd}&=\sqrt{\gamma_{3nd}}(\ket{24}\bra{8}). 
\end{aligned}
\end{equation}
Focusing on the dynamics of $\hat{H}_4$ first, we see that if we introduce the states $\ket{E_{\pm}}=\frac{1}{\sqrt{2}}\left(\ket{6}\pm\ket{10}\right)$ and assume that we are driving on resonance ($\Delta=0$), the Hamiltonian simplifies to 
\begin{eqnarray}
\hat{H}_4&=\Delta_3\ket{8}\bra{8}+\sqrt{2}g\left(\ket{E_+}\bra{E_+}-\ket{E_-}\bra{E_-}\right)\nonumber \\
&+\left(\Omega\left(\frac{1}{\sqrt{2}}\left(\ket{E_+}+\ket{E_-}\right)-\sqrt{9}\ket{8}\right)\bra{4}+h.c.\right). 
\end{eqnarray}
Since we are in the regime where $g\gg\Omega$, we can continue by eliminating the excited state manifold using the effective operator formalism as in the carving scheme. This gives the effective operators
\begin{eqnarray} \label{eq:effH41}
\hat{H}_{4eff}&\approx&-\tilde{\Delta}_4\ket{4}\bra{4} \\
\hat{L}_{4d}&\approx&\sqrt{\Gamma_{4d}}\ket{d}\bra{4} \\
\hat{L}_{4nd}&\approx&\sqrt{\Gamma_{4nd}}\ket{nd}\bra{4},
\end{eqnarray}
where
\begin{eqnarray} 
\tilde{\Delta}_4&\approx&\frac{9\Omega^2}{\Delta_3} \\
\Gamma_{4d}&\approx&\frac{9\Omega^2}{\Delta_3^2}\gamma_{3d}+\frac{\Omega^2}{2c}\gamma \\
\Gamma_{4nd}&\approx&\frac{9\Omega^2}{\Delta_3^2}\gamma_{3nd}+\frac{\Omega^2}{16c^2}\gamma_{nd},\label{eq:effH42}
\end{eqnarray}
in the limit $\Delta_3 \gg \gamma$ and $c\gg 1$ as in the carving protocol.

We now turn to the dynamics under the Hamiltonian $\hat{H}_3$ (still operating at resonance, i.e., $\Delta=0$). Assuming that $\Delta_3 \gg \sqrt{9}\Omega$ and that $\Delta_2\gg\sqrt{1/8}g$, we can adiabatically eliminate the coupling to the excited states $\ket{7}$ and $\ket{12}$. This results in the effective Hamiltonian
\begin{align}
\hat{H}_3\approx&-\frac{36\Delta_3\Omega^2}{4\Delta_3^2+\gamma^2}\ket{3}\bra{3}-\frac{1}{2}\frac{g^2\Delta_2}{4\Delta_2^2+\gamma^2}\ket{9}\bra{9}+\nonumber \\
&\left(\Omega\ket{5}\bra{3}+g(\sqrt{2}\ket{5}+\ket{11})\bra{9}+h.c.\right)
\end{align}
and approximate Lindblad operators
\begin{equation} \label{eq:lindbladg3}
\begin{aligned}
\hat{L}^{(3)}_{3d}&\approx\frac{2\sqrt{9\gamma_{3d}}}{2\Delta_3-i\gamma}\ket{21}\bra{3},\qquad \\
\hat{L}^{(3)}_{\tilde{e}1}&\approx\frac{\sqrt{\gamma}}{2}\frac{g}{2\Delta_2-i\gamma}\ket{27}\bra{9}, \qquad
\end{aligned}
\begin{aligned}
\hat{L}^{(3)}_{3nd}&\approx\frac{2\sqrt{9\gamma_{3nd}}}{2\Delta_3-i\gamma}\ket{23}\bra{3}, \\
 \hat{L}^{(3)}_{\tilde{e}2}&\approx\frac{\sqrt{\gamma}}{2}\frac{g}{2\Delta_2-i\gamma}\ket{28}\bra{9}.
\end{aligned}
\end{equation}
We now find the bright (has a cavity photon component) and dark states (no cavity photon component) of the Hamiltonian by diagonalizing the part that involves the cavity coupling between states $\ket{5},\ket{9}$, and $\ket{11}$. In the limit where $\Delta_2\gtrsim g$, we find that the bright states are
\begin{equation}
\ket{B_{\pm}}\approx\frac{1}{\sqrt{2}}\ket{9}\pm\left(\frac{1}{\sqrt{3}}\ket{5}+\frac{1}{\sqrt{6}}\ket{11}\right),
\end{equation}
while the dark state is $\ket{D}\approx\sqrt{\frac{2}{3}}\ket{11}-\frac{1}{\sqrt{3}}\ket{5}$. The Hamiltonian expressed in this basis is
\begin{align}
\hat{H}_3\approx & -\frac{36\Delta_3\Omega^2}{4\Delta_3^2+\gamma^2}\ket{3}\bra{3}+\Delta_+\ket{B_+}\bra{B_+}+\Delta_-\ket{B_-}\bra{B_-}\nonumber\\
&+\left(\left(\Omega_+\ket{B_+}+\Omega_-\ket{B_-}-\Omega/\sqrt{3}\right)\bra{3}+h.c.\right),
\end{align}
where $\Delta_{\pm}\approx \mp \sqrt{3}g$ and $\Omega_{\pm}\approx \Omega/\sqrt{3}$. It is seen that we are driving the transition to the dark state  on resonance, while the transitions to the bright states are off resonant. Consequently, we adiabatically eliminate the bright state dynamics, which results in an effective Hamiltonian and Lindblad operators
\begin{eqnarray} \label{eq:effH31}
\hat{H}_{3eff}&\approx&-\tilde{\Delta}_3\ket{3}\bra{3}, \\
\hat{L}_{3d}&\approx&\sqrt{\Gamma_{4d}}\ket{d}\bra{3}, \\
\hat{L}_{3nd}&\approx&\sqrt{\Gamma_{4nd}}\ket{nd}\bra{3}, \\
\hat{L}_{Dd}&\approx&\sqrt{\Gamma_{Dd}}\ket{d}\bra{D}, \\
\hat{L}_{Dnd}&\approx&\sqrt{\Gamma_{Dnd}}\ket{nd}\bra{D},
\end{eqnarray}
where
\begin{eqnarray} 
\tilde{\Delta}_3&\approx&\frac{9\Omega^2}{\Delta_3}+\frac{1}{36}\frac{\Omega^2}{\Delta_2^2}, \\
\Gamma_{3d}&\approx&\frac{9\Omega^2}{\Delta_3^2}\gamma_{3d}+\frac{2\Omega^2}{9c}\gamma, \\
\Gamma_{3nd}&\approx&\frac{9\Omega^2}{\Delta_3^2}\gamma_{3nd},\\
\Gamma_{Dd}&\approx&\frac{1}{3}\gamma_d+\frac{2}{3}\gamma, \\
\Gamma_{Dnd}&\approx&\frac{1/3}{\gamma_{nd}}.\label{eq:effH32}
\end{eqnarray}
Here the Lindblad operator $\hat{L}_{Dd}$ ($\hat{L}_{Dnd}$) describes decay from the dark state to detectable (non-detectable) states. It is seen that if the initial state is the ground state $\ket{3}$, a driving time of $t_g\approx\sqrt{3}\pi/\Omega$, will drive the population fully to the dark state and back again resulting in $\pi$-phase shift of the ground state up to a phase of $\tilde{\Delta}_3t_g$.

Using the effective Hamiltonians and Lindblad operators from Eqns.~(\ref{eq:effH41})-(\ref{eq:effH42}) and Eqns.~(\ref{eq:effH31})-(\ref{eq:effH32}), we find that starting from an initial, unentangled state of $\frac{1}{2}\left(\ket{0,0}+\ket{0,1}+\ket{1,0}+\ket{1,1}\right)$, the final state following a successful CZ gate (i.e., no detectable errors) will be
\begin{align}
\rho&\approx\frac{1}{P_g}(\ket{\psi_g}\bra{\psi_g}+\frac{1}{4}(\frac{\Gamma_{4nd}}{\Gamma_4}(1-e^{-\Gamma_4t})\\
&+\frac{\tilde{\Gamma}_{3nd}}{\tilde{\Gamma}_3}(1-e^{-\tilde{\Gamma}_3t}))\ket{nd}\bra{nd}),
\end{align}
where 
\begin{align}
\ket{\psi_g}&=\frac{1}{2}(\ket{0,0}+\ket{0,1}\\
&+e^{(i\tilde{\Delta}_3-\tilde{\Gamma}_3)/2)t_g}\ket{1,0}+e^{(i\tilde{\Delta}_4-\Gamma_4/2)t_g}\ket{1,1}),
\end{align}
with $\Gamma_4=\Gamma_{4d}+\Gamma_{4nd}$, $\tilde{\Gamma}_{3d}=(\Gamma_{3d}+\Gamma_{Dd})/2$, $\tilde{\Gamma}_{3nd}=(\Gamma_{3nd}+\Gamma_{Dnd})/2$, and $\tilde{\Gamma}_{3}=\tilde{\Gamma}_{3d}+\tilde{\Gamma}_{3nd}$. The success probability (i.e., probability not to detect the atoms in an error state) is
\begin{equation}
P_g\approx1-\frac{1}{4}\left(\frac{\Gamma_{4d}}{\Gamma_4}\left(1-e^{-\Gamma_4 t}\right)+\frac{\tilde{\Gamma}_{3d}}{\tilde{\Gamma}_3}\left(1-e^{-\tilde{\Gamma}_3 t}\right)\right).
\end{equation}
The fidelity with the target Bell state of $(\ket{0,0}+\ket{0,1}-\ket{1,0}+\ket{1,1})/2$ is 
\begin{equation}
F\approx\frac{1}{16P_g}\left|2+e^{(i\tilde{\Delta}_3-\tilde{\Gamma}_3)/2)t_g}+e^{(i\tilde{\Delta}_4-\Gamma_4/2)t_g}\right|^2. 
\end{equation}
Notably, this target Bell state is however not the optimal one due to the phase accumulation. This phase accumulation can be corrected using a local shift on one of the atoms. 

\clearpage

\end{document}